\documentclass[11pt]{article}

\usepackage[margin=1.1in]{geometry}
\usepackage{amsmath,amssymb}
\usepackage{booktabs}
\usepackage{graphicx}
\usepackage{microtype}
\usepackage[hidelinks]{hyperref}
\usepackage[numbers,sort&compress]{natbib}
\usepackage{enumitem}

\newcommand{\dsr}{\mathrm{DSR}}
\newcommand{\pbo}{\mathrm{PBO}}
\newcommand{\spa}{\mathrm{SPA}}
\newcommand{\mintrl}{\mathrm{MinTRL}}

\title{Equity Strategy Backtesting: Luck or Edge?\\
The MinervaScore as a Statistical Robustness Grade}

\author{
M.~L.~Santoni \quad V.~Jouanne \quad M.~L.~Scullin \\[4pt]
\normalsize Minerva (Minerva1.com) \\[2pt]
\normalsize \texttt{marialaurasantoni@gmail.com} \quad \texttt{info@minerva1.com}
}

\date{August 2026}

\begin{document}
\maketitle

\begin{abstract}
Backtests of trading strategies are often selected after many parameter
trials. A strong historical result can therefore reflect search luck rather
than a persistent signal. Standard summaries such as return, Sharpe ratio,
and drawdown do not record how many candidates were tried, whether the
selected rule survives out-of-sample validation, or whether the available
history is long enough to support the result.

This paper describes the MinervaScore, a post-selection robustness grade
for trading strategies. The score combines four established validation
quantities -- Deflated Sharpe Ratio, Probability of Backtest Overfitting,
Superior Predictive Ability, and Minimum Track Record Length -- with a
regime-stability diagnostic. These components are converted into signed
margins from their admissibility thresholds, aggregated into a raw score,
and then mapped to a 0--100 display. The display is tied to a binary
Robustness Seal: scores of 80 or higher are shown only when all five gates
pass. The calibration uses 359,062 production backtest records. The score
is intended to rank statistical support, not to estimate the probability of
future profit.

In synthetic markets with known ground truth, the MinervaScore separates
true signal from lucky backtest outcomes, with an AUROC of 0.989 at the
headline difficulty. Its improvement over the GT-Score proxy and the
gates-passed baseline is modest, and it remains close to the corrected
DSR-alone baseline. In a pre-registered test on unseen real-market data,
the score showed no significant forward relationship in a population with
limited surviving edge (Spearman $\rho_s = 0.013$, one-sided permutation
$p = 0.40$). We therefore present the MinervaScore as an auditable
validation and reporting layer, rather than as evidence of demonstrated
real-market predictability.

\end{abstract}

\section{Introduction}\label{sec:intro}

An investor optimizing a trading rule over ten years of data might test ten thousand parameter configurations and keep the one with the highest Sharpe ratio, the standard measure of return per unit of risk. That result is the maximum of ten thousand draws. Pure noise can also produce a high maximum, so a strong backtest says little by itself about whether the rule has an edge. A backtest reports an equity curve, the running total of simulated profit over the test period, while ignoring important questions such as how many combinations were tested, whether the winner held up across subperiods, and how much history supports it.

Three effects make a backtest a poor guide to future performance. First,
selection inflates the Sharpe ratio of the survivor as the number of trials
it beats increases \citep{bailey2014dsr}. Second, a rule tuned on one
stretch of history, the \emph{in-sample} period, often falls below the
median of its own trial cohort on data it has never seen, the
\emph{out-of-sample} period; the Probability of Backtest Overfitting
\citep{bailey2017pbo} measures this effect. Third, financial returns
violate the assumptions behind simple thresholds: volatility changes over
time, successive returns can be autocorrelated, and extreme moves are more
common than a Gaussian model implies \citep{cont2001stylized}. A threshold
calibrated under iid Gaussian assumptions can therefore be too permissive.

The literature provides several tools for addressing these problems:
the deflated Sharpe ratio, minimum track-record length, the SPA test, the
Model Confidence Set, FDR control, HAC-corrected Sharpe ratios, and
combinatorial purged cross-validation
\citep{bailey2014dsr, bailey2012sharpe, hansen2005spa, hansen2011mcs,
benjamini1995fdr, lo2002sharpe, lopezdeprado2018afml}. In practice,
however, applying them leaves the trader with heterogeneous statistics and
a pass/fail verdict rather than a single interpretable robustness grade.

Our aim is to turn the five validation quantities used by the
MinervaScore into a single value. This cannot be done by a simple average: four comfortable passes should not
be allowed to offset one disqualifying failure, and a display meant to show
admissibility has to preserve that logic. This paper documents MinervaScore, the aggregation and display layer running in production at Minerva,\footnote{\url{https://minerva1.com}} along with the audit we performed on it.

The MinervaScore is a validator applied after the search. It is computed on
the strategy already selected by the search, and it grades that strategy
from 0 to 100 based on the evidence behind it. Five validation gates compose the grade. Four are based on established
statistical procedures: the deflated Sharpe ratio, the probability of
backtest overfitting, superior predictive ability, and minimum track-record
length. Each gate controls a different failure mode. The DSR gate addresses
selection luck: it asks whether the selected Sharpe ratio remains unusual
after accounting for the number of strategies or parameter configurations
tried before that strategy was kept. The PBO gate addresses selection
instability: it asks whether the in-sample winner tends to fall below the
median of its own candidate cohort out of sample. The SPA gate addresses
benchmark luck: it asks whether the strategy remains superior to the
benchmark after accounting for dependence and repeated comparison. The
MinTRL gate addresses evidence length: it asks whether the available track
record is long enough to support the observed Sharpe ratio at the chosen
confidence level. The fifth gate is our own regime-stability diagnostic. It
addresses regime dependence by asking whether the evidence is concentrated
in a single market regime rather than distributed across materially
different conditions. Each gate compares the observed result with a
threshold derived from its corresponding validation quantity. When a
strategy clears all five gates, it is declared a pass, and we measure how
far it sits from each threshold, combining the resulting margins into the
continuous grade. A Seal pass therefore has a stated statistical interpretation defined by the
five gate thresholds. The displayed score
summarizes relative standing within the pass/fail verdict, and the
construction explicitly accounts for how many strategies were tried before
the selected one was kept.

The closest published composite that a practitioner is likely to encounter is the GT-Score \citep{sheppert2026gtscore}. The two should not be confused. The GT-Score is an empirical construction: a hand-designed formula that multiplies descriptors of an equity curve (mean return, an excess-return term, $R^2$, downside deviation), constructed to be maximized during the search that produces a strategy. Its value has no null model and no attached significance interpretation, and, most important, it carries no multiple-testing correction, since a curve-quality objective never observes how many strategies were tried. The two scores are applied at different points in the pipeline (\S\ref{sec:pipeline}) and answer different questions. The difference is in nature rather than measured accuracy: \S\ref{sec:synthetic} shows that the two discriminate genuine edge about equally well, so the case for the statistically grounded score rests on its threshold
interpretation, its multiple-testing correction, and the auditability of its
components. The two roles are complementary. A search objective finds candidates and a validator certifies them; the same measure should not do both, because a validator is trustworthy only when it was not the target of the search (\S\ref{sec:discussion}).

Table~\ref{tab:compare} summarizes the empirical evaluation performed in the synthetic environment described in \S\ref{sec:synthetic}, where the ground truth is known. Performance is assessed using the area under the receiver operating characteristic curve (AUROC), which measures the probability that a randomly selected strategy with a genuine edge is ranked above one without. An AUROC of 0.5 corresponds to random ranking, while a value of 1 indicates perfect discrimination \citep{hanley1982roc, fawcett2006roc}.
At the \emph{headline difficulty} (a default configuration used for the
primary results; the full grid over history length and edge strength is
defined in \S\ref{sec:synth-sweep}), the MinervaScore scores slightly above
the GT-Score proxy \citep{sheppert2026gtscore}, with the improvement
supported by a positive bootstrap confidence interval. Across the
pre-specified difficulty grid, it is the most stable composite score.
The corrected DSR alone remains the strongest single-signal baseline in
parts of the grid and accounts for most of the discriminative performance
after the null benchmark correction (\S\ref{sec:synth-srvariance}). The full composite, however, provides additional information by capturing
overfitting consistency, family-wise luck, track-record sufficiency, and
sensitivity to regime changes, aspects that the DSR alone does not cover.
It also preserves the multiple-testing correction carried by the DSR gate,
which is absent from curve-quality-based composites. Finally, the display
is aligned with the admissibility criteria: a score of 80 or higher is
shown only when all validation gates pass.

\begin{table}[t]
\centering
\small
\setlength{\tabcolsep}{5pt}
\begin{tabular}{lccccc}
\toprule
Scorer & Inputs & \shortstack{Multiple-testing\\correction} &
\shortstack{Verdict-consistent\\display} &
\shortstack{Genuine edge\\(AUROC)} & \shortstack{Worst-case\\regret}\\
\midrule
MinervaScore        & five gates   & yes & yes & \textbf{0.989} & \textbf{0.021}\\
DSR alone           & one gate     & yes & no  & 0.988 & 0.009\\
GT-Score (proxy)    & equity curve & no  & no  & 0.986 & 0.071\\
Gates passed        & five gates   & yes & no  & 0.960 & 0.138\\
\bottomrule
\end{tabular}
\caption{The MinervaScore against the baselines of \S\ref{sec:synthetic}, under the corrected DSR null benchmark. Each scorer is evaluated on a single task: discriminating genuine edge, which is the validator's job. The GT-Score, a search objective (\S\ref{sec:pipeline}), appears here as a familiar continuous benchmark measured on a task it was not designed for, and not as a competing validator. ``Genuine edge (AUROC)'' is the probability that a strategy with real edge outranks one without, at the headline synthetic difficulty; worst-case regret is the largest per-cell AUROC deficit to the best scorer across the pre-specified difficulty grid of \S\ref{sec:synth-sweep}. MinervaScore scores slightly above the GT-Score proxy with a positive
bootstrap interval; the corrected DSR alone is the strongest single-signal baseline, with the
smallest regret overall, and the composite's case rests on battery coverage
and verdict consistency rather than raw discrimination.}
\label{tab:compare}
\end{table}

\textbf{Contributions.}
\emph{(1) A score that preserves the ranking of strategies when reported
validation values become saturated} (\S\ref{sec:construction}). In practice,
many strategies fail the Deflated Sharpe Ratio (DSR) gate
($\dsr < 0.95$), and in the calibration population 60--95\% of DSR values
are reported as exactly zero, depending on the strategy family. The reported
DSR value therefore cannot distinguish between strategies that all fail by
different amounts. To preserve this information, we use the underlying DSR
test statistic and apply the log-odds (logit) transformation to bounded
validation quantities before combining them into a single score. This
approach allows strategies that fail by a small margin to be distinguished
from those that fail by a larger one. The pass/fail decision remains exactly
the same; only the ranking of strategies within the same verdict is refined.
If the pre-$\Phi$ statistic required for the DSR is not available, no score
is assigned rather than relying on an approximation.
\emph{(2) A displayed score consistent with the verdict}
(\S\ref{sec:display}). The 0--100 score is designed so that a value of 80 or higher appears only if the strategy passes all five gates, while a score below 80 indicates that at least one validation gate failed. This guarantee is exact. The small discontinuity introduced at the threshold is explicitly quantified rather than ignored (\S\ref{sec:audit}).
\emph{(3) Validation using a known ground truth} (\S\ref{sec:synthetic}). We evaluate the score in a simulated market where the true quality of each strategy is known, allowing us to distinguish strategies with genuine edge from those that only appear successful by chance. The proposed score remains competitive with the tested baselines and is
more stable than the other composite scores as the task becomes more
difficult (Table~\ref{tab:compare}). We also analyze the contribution of each component and examine how both the overall score and the individual gates are affected by the search size and the available history length. Finally, we use the same simulation framework to assess proposed modifications to the score, rejecting one because it inflated performance estimates and supporting another, a corrected luck benchmark, based on the empirical results (\S\ref{sec:synth-srvariance}).
\emph{(4) A pre-registered evaluation on real market data} (\S\ref{sec:realmarket}). We pre-specified the analysis and applied it once to a real-market dataset that had not been used during score development. In this dataset, which contains little evidence of genuine edge, we find no evidence that the score predicts out-of-sample outcomes. We report these null results in full and explain why they replace the conclusions of an earlier study, whose null findings can be attributed to a design flaw.
\emph{(5) A public audit based on 359{,}062 real results} (\S\ref{sec:audit}). The audit shows that no strategy saturates the top of the scale, the pass/fail guarantee holds without exceptions, and the ranking remains largely unchanged when all tuning parameters are varied by $\pm 20\%$. We also quantify the contribution of each of the five gates to the final score.

\textbf{Scope.} We do not claim that the score has demonstrated forward predictive power in real markets. The pre-registered sealed-window study produced null results, which are reported in \S\ref{sec:rm-primary}. Likewise, we do not interpret $\Phi(S-c)$ as a calibrated probability. Instead, we use it as a ranking measure, and \S\ref{sec:synth-calibration} evaluates how closely it approximates a calibrated probability. Finally, the continuous scoring framework is used only to rank strategies within the same pass/fail category and never across the pass/fail boundary.

\section{The Search Pipeline}\label{sec:pipeline}

A trading strategy is a set of trading rules together with the parameter
values used by those rules. For example, a moving-average crossover strategy
specifies buying when a short moving average crosses above a long one and
selling when it crosses back, with parameters such as the two averaging
windows, a minimum crossover distance, and a stop-loss level. A \emph{model}
is a family of strategies that share the same rule structure but differ in
their parameter values. For instance, a Bollinger-band mean-reversion
strategy, a Kalman-filter-based strategy, and a moving-average crossover
strategy are different models, each with its own parameter space. The
parameter search is therefore determined by the chosen model.

\subsection{Where optimization enters}
The user first selects a trading model, an instrument (or a universe of instruments), a historical period, and a bar interval. The Minerva platform then searches for parameter values that optimize the chosen objective. Each candidate parameter set is evaluated by running a full backtest, simulating the strategy bar by bar over the selected period using the platform's execution and transaction cost model, and computing a performance metric from the resulting equity curve. For a single instrument, the objective is typically the annualized Sharpe ratio. For a universe of instruments, it can be an aggregate measure such as the mean, median, or worst-case Sharpe ratio, or the Sharpe ratio adjusted for maximum drawdown. Because the parameter space is large and each backtest is computationally expensive, an exhaustive search is not feasible.

The way this objective is estimated also depends on the evaluation procedure. In the simplest setting, each candidate is evaluated on a single in-sample backtest. Alternatively, the platform can use walk-forward analysis or cross-validation, where performance is measured \emph{out of sample}. The remainder of this paper focuses on these out-of-sample evaluation procedures.

\subsection{The genetic optimizer}

The parameter search uses a genetic algorithm \citep{holland1975adaptation, goldberg1989genetic}. It begins with a randomly generated population of parameter sets (50 by default), each evaluated using the chosen objective. The population is then evolved for a fixed number of generations (20 by default). At each generation, the best-performing parameter sets are retained unchanged (elitism), while the remaining candidates are generated through tournament selection \citep{miller1995genetic}, uniform crossover, and Gaussian mutation. Tournament selection repeatedly compares a small random subset of candidates and selects the highest-scoring one as a parent. Uniform crossover combines the parameters of two parents with probability 0.7, while Gaussian mutation perturbs individual parameters with probability 0.1. Over successive generations, the search gradually concentrates on the most promising regions of the parameter space, and the highest-scoring parameter set found during the search is returned as the \emph{winner}.

\subsection{Optimizing out-of-sample: the genetic algorithm inside CPCV}
\label{sec:pipeline-cpcv}

The objective optimized by the genetic algorithm strongly influences the amount of overfitting produced during the search. In the walk-forward and cross-validation modes, optimization is performed using combinatorial purged cross-validation (CPCV) \citep{lopezdeprado2018afml}. The historical data are divided into multiple overlapping training and testing splits, with a \emph{purge} around each boundary and an \emph{embargo} immediately afterward to prevent information leakage from the test set into the training data.

Within each split, the genetic algorithm is trained only on the training data to identify the best parameter values. These parameters are then applied, without modification, to the corresponding held-out test data, and only the out-of-sample performance is used to evaluate the candidate. The overall fitness of a parameter configuration is obtained by aggregating its performance across all CPCV splits, so the search favors configurations that generalize well rather than those that overfit a single training sample.

Optimizing this objective instead of a single in-sample backtest has several advantages. First, it reduces overfitting during the search, since parameter configurations that perform well only on the training data receive poor out-of-sample scores and are naturally discarded. Second, it provides a more stable performance estimate by averaging results over many out-of-sample paths instead of relying on a single train/test split, making the search less sensitive to a favorable partition of the data \citep{arian2024backtest}. Finally, CPCV produces the candidate-by-fold performance matrix required by
the overfitting diagnostic, the Probability of Backtest Overfitting (PBO). When this matrix is fully available, PBO is computed using Combinatorially
Symmetric Cross-Validation (CSCV), described in \S\ref{sec:related}. When
the full matrix is unavailable, the platform reports a labeled
rank-correlation proxy rather than presenting the value as the true CSCV
estimator. CPCV therefore plays two roles: it defines the objective
optimized during the search and provides the input for one of the five
validation gates. It is also the evaluation protocol used in the synthetic
ground-truth study described in \S\ref{sec:synthetic}.

\subsection{Why the search makes a validator necessary}

Searching out of sample reduces overfitting but does not eliminate it. During a genetic search, hundreds or thousands of parameter configurations are evaluated, and the one with the highest score is selected. Even if each configuration is assessed on held-out data, the selected strategy is still the best among many noisy estimates, which introduces an upward selection bias. This bias increases as the search becomes more extensive, for example, by exploring more generations, wider parameter ranges, or more instruments.

Cross-validation therefore provides an out-of-sample estimate of performance, but it does not guarantee that the selected strategy will remain successful after re-selection, over a longer history, or under different market conditions. The optimizer's role is to identify the parameter configuration that generalizes best according to the chosen objective, not to determine whether its apparent edge is genuine or simply the result of an extensive search.

This is the role of the MinervaScore. It is computed only after the search has finished and is applied to the selected strategy rather than used to guide the optimization. Keeping these two stages separate is essential: a validation metric is informative only if it is evaluated on a strategy that was not selected to maximize it (\S\ref{sec:discussion}). A high MinervaScore therefore indicates stronger statistical support after
accounting for search effort, overfitting risk, and sensitivity to market
regime.

\section{Related Work and Terminology}\label{sec:related}

\paragraph{Multiple-testing-aware Sharpe ratios.}
The Deflated Sharpe Ratio (DSR) \citep{bailey2014dsr} adjusts the observed Sharpe ratio for two sources of bias: selection bias arising from multiple testing, via the expected maximum under the null over $N$ trials, and non-Gaussian return distributions. The latter uses the i.i.d.\ variance correction of \citet{mertens2002comments}, which accounts for skewness and kurtosis but not serial correlation. Serial correlation is handled separately through the heteroskedasticity and autocorrelation consistent (HAC) correction of \citet{lo2002sharpe, newey1987simple}, which we report as an audit metric rather than incorporate into the DSR. \citet{lo2002sharpe} also derived the time aggregation of the Sharpe estimator's sampling variance, which we use as the null benchmark for the DSR gate (\S\ref{sec:synth-srvariance}).

In our implementation, the multiple-testing correction is based on the \emph{effective} number of independent trials rather than the total number of parameter configurations evaluated. This effective number is estimated by decorrelating the searched parameter vectors using an eigenvalue-based approach, so that highly correlated parameter configurations are not counted as independent searches.

Two related length measures should be distinguished. Our fourth validation gate is the Minimum Track Record Length (MinTRL) proposed by \citet{bailey2012sharpe}. It gives the minimum track length required for an observed Sharpe ratio to be statistically different from zero with confidence $1-\alpha$:

\begin{equation}
\mintrl \;=\; 1 + \Big(1 - \gamma_3\, SR + \tfrac{\gamma_4 - 1}{4}\, SR^2\Big)
\left(\frac{\Phi^{-1}(1-\alpha)}{SR}\right)^{\!2},
\label{eq:mintrl}
\end{equation}
where $SR$ is the per-bar Sharpe ratio, $\gamma_3$ is the skewness, $\gamma_4$ is the raw kurtosis, and $\alpha = 0.05$. The coefficient $(\gamma_4-1)/4$ can equivalently be written $(\kappa+2)/4$, where $\kappa$ is the excess kurtosis.

MinTRL depends only on the properties of a single strategy and is independent of the number of trials $N$. It should therefore not be confused with the Minimum Backtest Length (MinBTL, $\mathrm{MinBTL} \approx 2\ln N / E[\max]^2$; \citealp{harvey2016cross, lopezdeprado2018afml}), which explicitly accounts for multiple testing. Another related approach is the non-linear haircut of \citet{harvey2015backtesting}, a direct alternative to the DSR correction that we leave for future work.

\paragraph{Family-wise inference.}
The Superior Predictive Ability (SPA) test \citep{hansen2005spa} extends White's Reality Check \citep{white2000reality} by providing a less conservative correction for multiple testing. Other methods, such as the Romano--Wolf StepM procedure \citep{romano2005stepwise} and the Model Confidence Set (MCS) \citep{hansen2011mcs}, identify the models that remain significant after rejecting the overall family of hypotheses. False Discovery Rate (FDR) procedures \citep{benjamini1995fdr, benjamini2001fdr} instead control the expected proportion of false discoveries rather than the family-wise error rate. In our framework, the SPA test is used as one of the validation gates, while the remaining methods are reported as audit metrics.

\paragraph{Cross-validation in finance.}
Combinatorial Purged Cross-Validation (CPCV) \citep{lopezdeprado2018afml} extends standard cross-validation by applying purging and embargoing around test windows while maximizing the number of out-of-sample evaluation paths that can be obtained from a finite historical dataset. With $N$ groups and $k$ test groups, CPCV generates $\binom{N}{k}$ train/test splits, corresponding to $\varphi[N,k] = \binom{N}{k}\, k / N = \binom{N-1}{k-1}$ distinct backtest paths. For example, $N=6$ and $k=2$ produce 15 train/test splits but only 5 distinct paths. This distinction is important whenever the number of evaluation folds is interpreted as an effective sample size.

A related procedure, Combinatorially Symmetric Cross-Validation (CSCV) \citep{bailey2017pbo}, provides the basis for the Probability of Backtest Overfitting (PBO) estimate used in this paper. The candidate-by-fold performance matrix is divided into all symmetric in-sample and out-of-sample combinations, and PBO measures the fraction of cases in which the in-sample winner ranks below the median out-of-sample performance.

\citet{arian2024backtest} show that CPCV-based validation is effective at distinguishing genuine predictive ability from selection bias in synthetic experiments, motivating the evaluation framework used in \S\ref{sec:synthetic}. \citet{yin2026implementation} show that backtesting engines produce similar results when transaction costs are ignored but may diverge significantly once costs are included. This motivates explicitly storing the cost-model configuration with every reported result.

\paragraph{Composite scores.}
The closest published composite benchmark is the GT-Score \citep{sheppert2026gtscore}, a multiplicative objective designed to be \emph{maximized during the search}:
\[
\text{GT} = m \cdot \ln z \cdot r^2 / s_d,
\]
where $m$ is the mean per-trade return, $z$ is a $t$-style excess-return term, $r^2$ measures equity-curve fit, and $s_d$ is the downside deviation. The GT-Score is a heuristic performance measure that combines several equity-curve characteristics but does not attach a null model or statistical interpretation to its value. In contrast, the MinervaScore combines validation quantities drawn from the statistical backtesting literature with an explicit regime-stability
diagnostic, and ties the pass/fail decision to fixed gate thresholds.

The two methods are designed for different stages of the pipeline: the GT-Score is used as an optimization objective during the search, whereas the MinervaScore is applied after the search as a validation and reporting measure. In addition, the GT-Score does not include an explicit correction for multiple testing. We therefore use it in \S\ref{sec:synthetic} as a familiar continuous
benchmark, and Table~\ref{tab:compare} reports the comparison.

\paragraph{Aggregation lineage.}
The aggregation in Eq.~\eqref{eq:aggregate} follows the weighted inverse-normal combination proposed by \citet{liptak1958combination} (with the unweighted version introduced by \citealp{stouffer1949american}) and uses a denominator adjusted for dependence, in the spirit of \citet{hartung1999note}. The weighting rationale is further supported by \citet{whitlock2005combining}. Brown's method \citep{brown1975method} and related empirical extensions \citep{poole2016combining, cinar2022poolr} instead modify Fisher's combination statistic on the $\chi^2$ scale and represent a different aggregation framework; we mention them only as alternative approaches that we did not adopt.

Since our margins are descriptive coordinates rather than $p$-values from tests of a common null hypothesis (\S\ref{sec:margins}), we use this literature to justify the algebraic form of the aggregation, not to claim an inferential interpretation.

\section{The MinervaScore Construction}\label{sec:construction}

\begin{figure}[t]
\centering
\includegraphics[width=\linewidth]{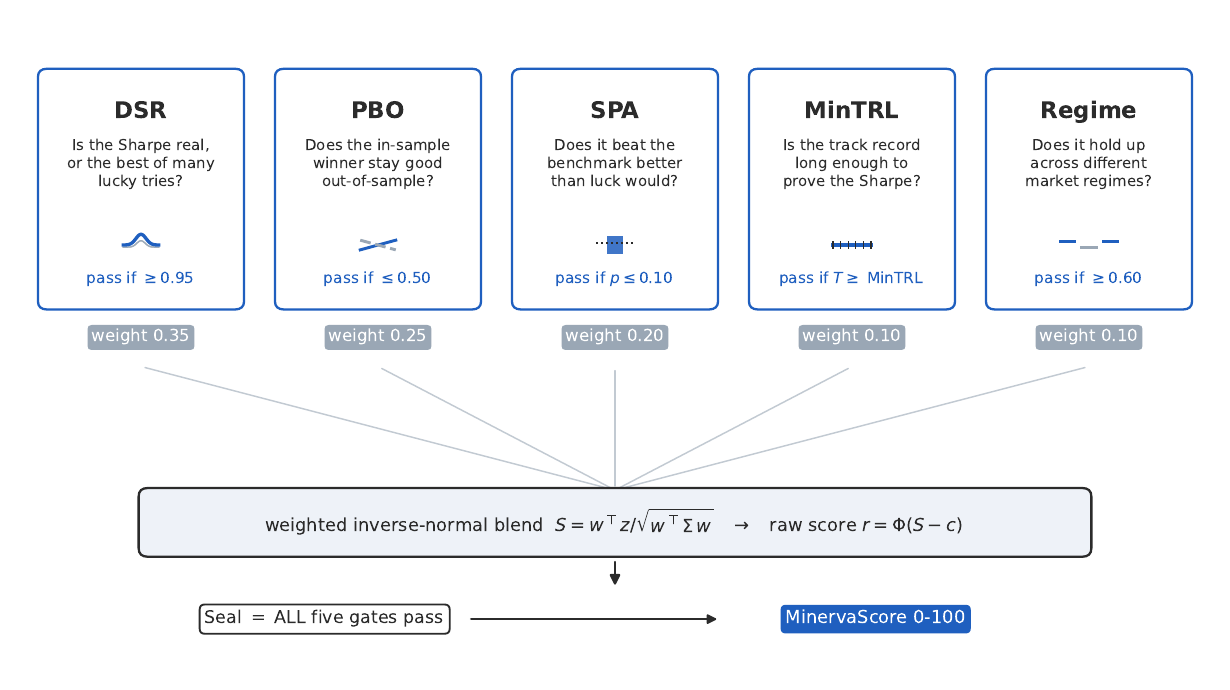}
\caption{The five validation gates and their combination into the MinervaScore. Each gate evaluates a different aspect of a candidate strategy and is passed only if its corresponding threshold is met. A strategy receives the Robustness Seal only if it passes all five gates. To compute the continuous score, each gate is first converted into a signed margin relative to its threshold (\S\ref{sec:margins}). These margins are then combined using a weighted inverse-normal aggregation (\S\ref{sec:aggregation}) to produce a raw score ($r \in [0,1]$). Finally, $r$ is transformed into the displayed MinervaScore on a 0--100 scale (\S\ref{sec:display}), with scores between 80 and 100 reserved for strategies that obtain the Robustness Seal.}
\label{fig:schematic}
\end{figure}

The MinervaScore is computed in three stages, illustrated in Fig.~\ref{fig:schematic}. First, five validation gates determine whether a strategy passes or fails (\S\ref{sec:seal}). Second, a continuous raw score is computed by combining the signed margins of each gate relative to its threshold (\S\ref{sec:margins}--\S\ref{sec:offset}). Finally, the raw score is mapped to a displayed value on a 0--100 scale that remains consistent with the pass/fail verdict (\S\ref{sec:display}).

\subsection{The Seal and the evidence floor}
\label{sec:seal}

A strategy receives the Robustness Seal if and only if it passes all five validation gates:
\begin{equation}
\mathrm{Seal} \;=\; \mathbf{1}\!\left[\,
\dsr \ge \tau_{\dsr} \;\wedge\; \pbo \le \tau_{\pbo} \;\wedge\;
\spa \le \tau_{\spa} \;\wedge\; T \ge \mintrl \;\wedge\; \rho \ge \tau_{\rho}
\,\right],
\label{eq:seal}
\end{equation}
where $\tau_{\dsr}=0.95$, $\tau_{\pbo}=0.50$, $\tau_{\spa}=0.10$, $\tau_{\rho}=0.60$, and the MinTRL threshold is defined by \citet{bailey2012sharpe}. The Seal is therefore the logical intersection (AND) of the five validation criteria.

The first four gates correspond to established methods introduced in \S\ref{sec:related}. The fifth gate, the regime composite $\rho$, is specific to our framework and is defined as
\begin{equation}
\rho \;=\; 0.5\,p_{+} \;+\; 0.3 \cdot \mathrm{clip}_{[0,1]}\!\left(
1 - \frac{s_{\mathrm{SR}}}{\sigma_{\mathrm{ref}}}\right) \;+\;
0.2 \cdot \mathrm{logistic}\!\left(SR_{\min}\right),
\label{eq:regime}
\end{equation}
where the quantities are computed over the validation windows of a run.
Here, $p_{+}$ is the fraction of windows with a positive Sharpe ratio,
$s_{\mathrm{SR}}$ is the standard deviation of the per-window Sharpe ratios,
$SR_{\min}$ is the minimum Sharpe ratio across the windows,
$\sigma_{\mathrm{ref}} = 2$ is a fixed reference scale, and
$\mathrm{logistic}(x) = 1/(1+\exp(-x))$. The final value is clipped to
$[0,1]$, and the three component weights are chosen heuristically. Their influence on the final score is analyzed in \S\ref{sec:audit}, while \S\ref{sec:discussion} discusses this gate as the most heuristic component of the proposed framework.

In addition to the five validation gates, the platform records an
\emph{evidence floor}. Results with limited empirical support---for example,
too few trades, too few independent candidates, or too few validation
windows---are flagged separately as having insufficient evidence. This flag
is not a sixth validation gate and is not combined into the continuous
score. Instead, it qualifies the interpretation of the result: a strategy
with too little evidence should not be presented as statistically certified,
but it should also be distinguished from a strategy that has enough evidence
and fails one of the five validation gates.

\subsection{Gate margins on two scales}
\label{sec:margins}

Each validation gate contributes to the continuous score via its signed distance from the admissibility threshold, standardized by a cross-sectional dispersion $\sigma_k$ estimated once from a calibration population (\S\ref{sec:calibration}). The appropriate scale for measuring this distance depends on the distribution of the gate values.

On their original $[0,1]$ scale, several gate values are highly concentrated.
In our calibration population, the DSR is exactly $0.0$ for 60--95\% of the
records, depending on the strategy family. This occurs because the DSR is
defined as $\Phi(u)$, where $u$ is the deflated test statistic, and $\Phi$
maps strongly negative values ($u \lesssim -8$) to floating-point zero.
Long-history strategies frequently produce $u \in [-40,-15]$, causing a
large fraction of the population to collapse to the same value. Similar
boundary effects occur for PBO and SPA. In the earlier campaign used to
freeze the bounded-gate dispersions, PBO had a large point mass at exactly
$0$; in the current production population this mass remains material,
although lower. SPA also accumulates at its bootstrap lower bound. As a
result, meaningful dispersion cannot be estimated reliably on the raw
$[0,1]$ scale.

\begin{align}
z_{\dsr} &= \frac{u - \Phi^{-1}(\tau_{\dsr})}{\sigma_{\dsr}},
&u &= \frac{\widehat{SR} - SR_0}{\widehat{se}}
\quad(\text{so } \dsr = \Phi(u)),
\label{eq:dsr-margin}\\[2pt]
z_{k} &= \frac{\mathrm{logit}(\tau_k) - \mathrm{logit}(g_k)}{\sigma_k}
\quad (k \in \{\pbo, \spa\}),
&z_{\rho} &= \frac{\mathrm{logit}(\rho) - \mathrm{logit}(\tau_\rho)}{\sigma_\rho},
\label{eq:logit-margins}
\end{align}
where $\mathrm{logit}(p)=\log\!\big(p/(1-p)\big)$ is computed after clamping $p$ to $[\varepsilon,1-\varepsilon]$, with $\varepsilon=10^{-6}$, so that boundary values remain finite. The pre-$\Phi$ statistic $u$ is the natural coordinate for the DSR because it is the underlying test statistic, is unbounded, and preserves distinctions that are lost after applying $\Phi$. For example, strategies with $u=-15$ and $u=-40$ both have $\dsr=0.0$, even though they represent very different levels of evidence. In our implementation, $u$ is computed using the Lo null sampling variance of the Sharpe estimator as the deflation benchmark; the motivation for this choice is discussed in \S\ref{sec:synth-srvariance}. The logit transformation plays an analogous role for PBO, SPA, and the regime composite $\rho$. Since both $\Phi$ and the logit are strictly monotonic, these transformations preserve the sign of every margin and therefore leave the Seal unchanged. They affect only the continuous ranking of strategies within the same pass/fail category.

The MinTRL margin is unchanged from the original formulation. Because the raw quantity $(T-\mintrl)/\sigma_T$ grows without bound for long track records, it is mapped through a hyperbolic tangent:
\begin{equation}
z_{\mintrl} = \tanh\!\left(\frac{T - \mintrl}{\sigma_T}\right),
\qquad \sigma_T = \max(0.2\,\mintrl,\; 50).
\label{eq:mintrl-margin}
\end{equation}

These standardized margins are descriptive rather than inferential. Dividing by the cross-sectional dispersion produces a population-standardized measure of the distance from the threshold, not a variable that follows an $N(0,1)$ distribution under the null hypothesis. Consequently, we make no inferential claim for the individual $z_k$ values or for the aggregate score defined below.

\paragraph{Fail-closed on a missing statistic.}
The pre-$\Phi$ representation requires the statistic $u$ to be stored when the DSR is computed. It cannot be reconstructed from the reported DSR value because $\Phi^{-1}(0.0)$ is effectively clamped to $\Phi^{-1}(\varepsilon) \approx -4.75$, causing all strongly overfit strategies to collapse to the same value and reintroducing the saturation that the proposed construction is designed to avoid.

For this reason, the implementation fails closed. If a record contains a DSR value but not the corresponding pre-$\Phi$ statistic, no MinervaScore is produced and a machine-readable explanation is returned instead of an approximate score. Older records created before the statistic was stored therefore remain unscored until they are recomputed. This design preserves the integrity of the displayed score, in the same way as the display invariant described in \S\ref{sec:display}.

\subsection{Aggregation: correlation-adjusted weighted inverse-normal}
\label{sec:aggregation}

The five standardized margins are combined using the same aggregation as in the original raw-scale formulation:
\begin{equation}
S = \frac{w^{\top} z}{\sqrt{w^{\top} \Sigma_{\mathrm{eff}}\, w}},
\qquad w = (0.35,\, 0.25,\, 0.20,\, 0.10,\, 0.10),
\label{eq:aggregate}
\end{equation}
This is the weighted inverse-normal combination of \citet{liptak1958combination} (with the unweighted version due to \citealp{stouffer1949american}), using a dependence adjustment in the spirit of \citet{hartung1999note}; the argument for weighting follows \citet{whitlock2005combining}. The matrix $\Sigma_{\mathrm{eff}}$ is the cross-sectional correlation matrix of the standardized margins, estimated once on the calibration population (\S\ref{sec:calibration}) and shared across all strategies to ensure that scores remain comparable.

The denominator is a fixed normalization constant that places $S$ on an approximately unit scale. It is intended as a variance-stabilizing rescaling, not as an inferential correction for combining dependent tests. It would coincide with the cross-sectional standard deviation of $w^{\top}z$ only if every standardized margin had unit cross-sectional variance, which is not the case. The dispersions of the bounded gates are estimated after including the clamped values (\S\ref{sec:calibration}), the $\sigma_k$ are based on the interquartile range rather than the second moment, and $z_{\mintrl}$ is a bounded $\tanh$ transformation that does not use a corresponding $\sigma_k$. As a result, the normalization constant slightly overestimates the realized spread of $w^{\top}z$, compressing the values of $S$ without changing their ordering.

For the same reason, we do not adopt modern methods for combining dependent $p$-values \citep{brown1975method,poole2016combining,cinar2022poolr}, because these methods assume $p$-values from tests of a common null hypothesis, whereas our margins are descriptive quantities. When one or more gates are unavailable, the weights are renormalized over the remaining active gates. The weighting scheme reflects the prior importance assigned to each gate, with the DSR receiving the largest weight because it is the only gate that simultaneously accounts for selection bias and non-Gaussian returns. The practical effect of this weighting is evaluated in \S\ref{sec:audit}.

\subsection{The conservative offset}
\label{sec:offset}

The raw score is obtained by applying a one-sided offset $c \ge 0$ before the final normal CDF:
\begin{equation}
r \;=\; \Phi(S - c), \qquad c = 0.5 \text{ (operator default)}.
\label{eq:rawscore}
\end{equation}
The parameter $c$ acts as a conservative offset. It reduces the raw score by
$\Phi(S)-\Phi(S-c)=\int_{S-c}^{S}\varphi$,
with the largest reduction occurring around $S=c/2$ and progressively smaller effects in both tails. At the admissibility threshold ($S=0$), the reduction is $0.191$; at $S=3$ it decreases to $0.005$, and at $S=-3$ to $0.001$. Since most strategies in a typical population lie well below the admissibility threshold, the effect of the offset is negligible for clear failures. Instead, it primarily reduces borderline and potentially inflated scores.

Because $c$ is a constant shift inside the monotonic mapping $\Phi$, it does not affect the ordering of strategies. The induced ranking is therefore exactly invariant to the choice of $c$, as shown analytically and verified empirically in \S\ref{sec:audit}.

\subsection{Why not a Student-$t$ map}
\label{sec:student-t}

An alternative to \eqref{eq:rawscore} would be to replace $\Phi$ with a symmetric Student-$t$ CDF, motivated by the heavy tails commonly observed in financial returns \citep{cont2001stylized}. We do not adopt this approach. Unlike the original raw-scale construction, the aggregate $S$ computed from the two-scale margins is \emph{leptokurtic} on the calibration population (excess kurtosis $+2.77$), rather than platykurtic, because the unbounded pre-$\Phi$ DSR margin restores genuine tail behavior.

The choice of $\Phi$ is based on two considerations. First, a symmetric Student-$t$ CDF compresses both tails toward $\tfrac12$. In a population where most strategies fail, this would increase the displayed scores of poor strategies: for example, $S=-3$ maps to $0.0013$ under $\Phi$ but to $0.020$ under $t_4$. This is the opposite of the conservative behavior we seek, and exactly what the offset $c$ is intended to avoid. Second, the effect of heavy tails is already addressed within the individual validation gates: DSR accounts for skewness and kurtosis through the Mertens variance adjustment \citep{bailey2014dsr,mertens2002comments}, SPA uses a block bootstrap, and PBO and $\rho$ are based on rank- and bounded-scale statistics. Applying another heavy-tail correction at the display stage would therefore amount to correcting for the same effect twice.

Other monotone transformations could be studied, but each would constitute
a separate calibration choice. The Robustness Seal, which carries the
validation decision, would remain unchanged.

\subsection{Seal-consistent two-band display}
\label{sec:display}
The raw score in \eqref{eq:rawscore} does not by itself guarantee that a strategy that passes the Seal is ranked above every failing one. In a single continuous aggregation, four strong gates can compensate for one mandatory failure. The displayed score addresses this by conditioning on the Seal verdict. For failing strategies, it represents the \emph{percentile of relative standing} within the reference population of optimizer backtests.

Let $\widehat{F}$ denote the empirical CDF of the raw score computed from the frozen reference population (the 355{,}214 uncertified rows of the calibration ledger, stored as 81 quantile knots at 1.25\% intervals and evaluated by linear interpolation). The displayed score is then defined as
\begin{equation}
\mathrm{Display} = \left\lfloor D \right\rfloor, \qquad
D =
\begin{cases}
\;\min\!\left(80 + 20 \cdot \dfrac{r - r_0}{r^{\star} - r_0},\; 100\right),
& \text{Seal passed},\\[8pt]
\;\min\!\left(80 \cdot \widehat{F}(r),\; 80 - \delta\right),
& \text{Seal failed},
\end{cases}
\label{eq:display}
\end{equation}
where $r_0=\Phi(-c)$ is the minimum raw score achievable by an admissible strategy (all margins equal to zero, implying $S=0$), and $r^{\star}$ is the raw score of a theoretically perfect strategy, with every gate at its physical limit. With the current calibration constants, $r_0 \approx 0.3085$ and $r^{\star} \approx 0.9997$. The displayed score is reported as an integer, and $\delta=0.1$ is the default separation constant.

The cap on the failing branch is required because the reference CDF reaches $\widehat{F}(r)=1$ at its highest knot. Without the cap, the best failing strategy would receive a displayed score of exactly 80. Both the anchor values and the quantile knots are fixed for a given calibration version and are computed from the calibration ledger rather than hard-coded.

For failing strategies, the displayed value therefore has a direct interpretation as relative standing: apart from integer truncation, $\mathrm{Display}/0.8$ gives the percentage of all strategy configurations evaluated by the platform that the strategy ranks above in raw-score terms. The certified interval $[80,100]$ is reserved exclusively for strategies that obtain the Seal and is mapped linearly from the admissibility threshold to the theoretical maximum score. From \eqref{eq:display} it follows directly that
\begin{equation}
\mathrm{Display} \ge 80 \iff \text{Seal passed},
\label{eq:invariant}
\end{equation}
and this property is enforced by construction rather than established empirically. The analysis in \S\ref{sec:audit} therefore serves only to verify the implementation.

Finally, two observations explain why the transition at the certification threshold is small. First, the failing branch is strictly monotonic in $r$, so all ranking-based results reported in \S\ref{sec:synthetic} are independent of the display mapping. Second, in the production dataset the lowest raw score among sealed strategies is already 0.3915, corresponding to the 96th percentile of the reference distribution, or an unsealed display score of 77.5. Certification therefore changes the displayed score by only a few points near the threshold.

\begin{table}[t]
\centering
\small
\begin{tabular}{llrr c}
\toprule
MinervaScore & Raw score $r$ & \shortstack{Backtests\\in band} &
\shortstack{User results\\in band ($n=516$)} & Verdict\\
\midrule
0--9   & 0.0000--0.0006 & 44{,}402 & 9 (1.7\%)   & not sealed\\
10--19 & 0.0006--0.0022 & 44{,}402 & 2 (0.4\%)   & not sealed\\
20--29 & 0.0022--0.0049 & 44{,}401 & 7 (1.4\%)   & not sealed\\
30--39 & 0.0049--0.0087 & 44{,}402 & 12 (2.3\%)  & not sealed\\
40--49 & 0.0087--0.0148 & 44{,}402 & 30 (5.8\%)  & not sealed\\
50--59 & 0.0148--0.0279 & 44{,}401 & 105 (20.3\%) & not sealed\\
60--69 & 0.0279--0.0739 & 44{,}402 & 164 (31.8\%) & not sealed\\
70--79 & 0.0739--1.0000 & 44{,}402 & 186 (36.0\%) & not sealed\\
\textbf{80--100} & 0.3915--0.9999 & 3{,}848 & 1 (0.2\%) & \textbf{Sealed}\\
\bottomrule
\end{tabular}
\caption{The displayed MinervaScore on the production population. The failing bands (0--79) represent percentiles of the raw score with respect to the frozen reference population. By construction, each band contains $\approx 44{,}402$ reference backtests. The ``User results'' correspond to the best strategy from each optimization run ($n = 516$). Their median falls in the 60--69 band, and 88\% achieve a displayed score of at least 50, whereas the raw score alone would compress the same distribution into single-digit values. The certified band (80--100), containing 3{,}848 Sealed strategies, overlaps the raw-score range 70--79 because certification depends on passing all five validation gates rather than on a raw-score threshold. The transition is nevertheless almost seamless, since the lowest Sealed $r$ already lies at the 96th percentile of the reference distribution.}
\label{tab:display-bands}
\end{table}

In the production population, Sealed strategies receive displayed scores between 82 and 100, with a median of 91, while the highest score among failing strategies is 79. The distribution of the best result from each optimization run is summarized in Table~\ref{tab:display-bands}, with a median in the 60--69 band and 88\% of results scoring at least 50.

The platform distinguishes between the displayed score and the
underlying continuous score, and we follow the same convention throughout
this paper. The term \emph{MinervaScore} refers only to the displayed
0--100 value defined by Eq.~\eqref{eq:display}. The continuous value in $[0,1]$ is referred to as the \emph{raw score} $r$. In the API payload it
is stored as \texttt{score}, together with the displayed score.

The consequence of gating on the Seal verdict remains the same: the displayed band is determined entirely by the binary Seal, so the displayed score is discontinuous at the decision boundary, although the discontinuity is small because of the percentile mapping described above. All continuous quantities introduced in \S\ref{sec:margins}--\S\ref{sec:offset} therefore rank strategies only within the same band. The practical impact of this discontinuity is evaluated in \S\ref{sec:audit}.

As with all calibrated quantities in the framework, displayed scores are comparable only within the same calibration vintage. Updating the frozen quantile knots constitutes a versioned recalibration. Because the platform stores the raw score $r$ rather than the displayed value, previously computed results never need to be rewritten. Instead, the displayed score is recomputed at serving time from the stored (raw score, verdict) pair. As a result, strategies evaluated before a recalibration are displayed using the current scale, while their raw scores---and therefore their percentile positions---remain those of the calibration vintage in which they were originally computed. Only newly executed optimizations incorporate the updated deflation within the validation gates.

\section{Empirical Calibration}\label{sec:calibration}

The construction described in \S\ref{sec:margins} standardizes each gate margin using a dispersion constant $\sigma_k$, which represents the cross-sectional spread of that margin across a representative population of strategies. This defines the unit in which distance from the threshold is measured: a value of $z_k=1$ corresponds to one population spread beyond the threshold, on the natural scale of the corresponding gate.

The dispersion constants are estimated once from the production trial ledger, which contains \textbf{359{,}062} result rows with a persisted pre-$\Phi$ statistic, collected from \textbf{536} optimization runs spanning the platform's model families, bar intervals, and instruments. This dataset represents the operating population for which the constants are intended.

For the DSR, the dispersion is computed as the pooled $\mathrm{IQR}/1.349$ of the pre-$\Phi$ margins, where the interquartile range is divided by $\Phi^{-1}(0.75)-\Phi^{-1}(0.25)\approx1.349$. Under normality, this provides an estimate of the standard deviation while remaining robust to extreme values. The DSR deflation uses the Lo null sampling variance $1/\mathrm{years}$, as discussed in \S\ref{sec:synth-srvariance}.

\begin{table}[t]
\centering
\small
\begin{tabular}{lcc}
\toprule
Gate & Threshold $\tau_k$ & $\sigma_k$ (margin scale)\\
\midrule
DSR (pre-$\Phi$ $u$) & $\Phi^{-1}(0.95) = 1.645$ & 1.128\\
PBO (logit)          & $\mathrm{logit}(0.50) = 0$ & 10.028\\
SPA (logit)          & $\mathrm{logit}(0.10) = -2.197$ & 6.013\\
Regime $\rho$ (logit) & $\mathrm{logit}(0.60) = 0.405$ & 1.161\\
\bottomrule
\end{tabular}
\caption{Frozen per-gate dispersions on the margin scales of
\S\ref{sec:margins} (359{,}062 production rows; logit clamp
$\varepsilon = 10^{-6}$, identical to production). MinTRL carries no
$\sigma_k$: its tanh margin has the $\sigma_T$ scale built in
(Eq.~\ref{eq:mintrl-margin}).}
\label{tab:sigmas}
\end{table}

\paragraph{The bounded-gate dispersions are clamp-inclusive by design.}
A degenerate run-level $\pbo = 0$ (caused by collapsed CSCV combinatorics and occurring in 14.4\% of production rows) is mapped by the $\varepsilon$-clamp to a logit margin of $+13.8$. The large value of $\sigma_{\pbo}$ deliberately reduces the influence of this unreliable point mass, limiting its contribution to about $+1.4\sigma$ instead of allowing a clamping artifact to dominate the aggregate score. The same principle applies to the SPA bootstrap floor through $\sigma_{\spa}$.

The alternative is illustrated by the frozen calibration artifact. If the dispersion is estimated only from the clamp-free core of the population, it decreases to $\sigma_{\pbo} \approx 1.1$. In that case, a clamped observation contributes approximately $+12\sigma$, causing the raw score to saturate: 10\% of rows receive scores $\geq 0.98$, compared with only 0.002\% reaching $\geq 0.99$ under the frozen calibration constants. Given the choice between allowing an unreliable signal to dominate the aggregate or reducing its influence, the proposed construction adopts the latter.
\begin{table}[t]
\centering
\small
\begin{tabular}{lrrrrr}
\toprule
 & DSR & PBO & SPA & MinTRL & Regime\\
\midrule
DSR    & $1.00$ & $0.23$ & $0.32$ & $0.71$ & $0.25$\\
PBO    & $0.23$ & $1.00$ & $0.65$ & $0.48$ & $0.51$\\
SPA    & $0.32$ & $0.65$ & $1.00$ & $0.57$ & $0.60$\\
MinTRL & $0.71$ & $0.48$ & $0.57$ & $1.00$ & $0.55$\\
Regime & $0.25$ & $0.51$ & $0.60$ & $0.55$ & $1.00$\\
\bottomrule
\end{tabular}
\caption{Frozen $\Sigma_{\mathrm{eff}}$: winsorized-1\% Pearson
correlation of the standardized margins on the production population
(pairwise complete). The matrix is positive definite without projection
(smallest eigenvalue 0.21); the full-mask denominator is
$\sqrt{w^{\top}\Sigma w} = 0.753$.}
\label{tab:sigmaeff}
\end{table}

\paragraph{What the correlation structure shows.}
The DSR margin is positively correlated with all other validation margins (Table~\ref{tab:sigmaeff}), with the strongest association observed for MinTRL (0.71). This is expected, since both quantities reflect the amount of statistical evidence supporting a strategy, and strategies with stronger evidence tend to satisfy multiple validation criteria simultaneously. The normalization term in Eq.~\eqref{eq:aggregate} accounts for these correlations, preventing the aggregate score from rewarding the same underlying evidence more than once.

\begin{figure}[t]
\centering
\includegraphics[width=\linewidth]{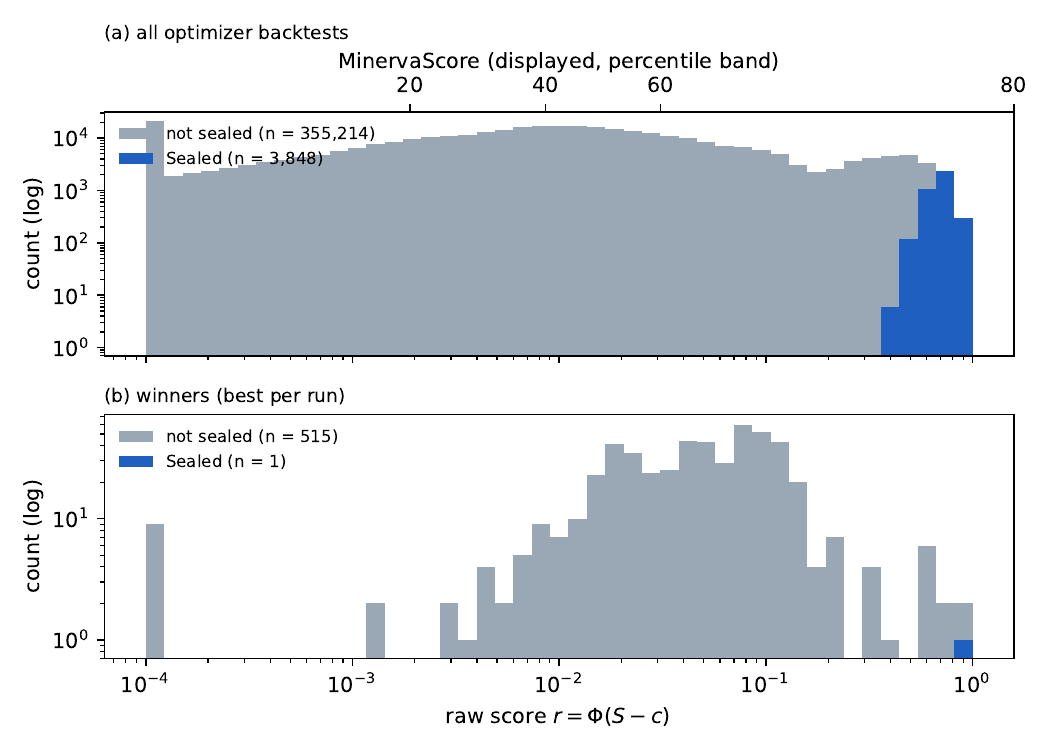}
\caption{Raw-score distribution by verdict, computed through the
production module under the frozen constants (log count). \textbf{(a)} all
359{,}062 scored backtests; \textbf{(b)} the 516 winners (best result per
optimization run). The top axis maps the raw score to the displayed
MinervaScore through the percentile scale of \S\ref{sec:display} (failing
band; Sealed rows, in blue, display 80--100 regardless of their raw
value). Nothing saturates at the ceiling, the failing mass is continuously
ordered, and the winners in (b) sit far above the full population.}
\label{fig:rawhist}
\end{figure}

\paragraph{Resulting distribution.}
Figure~\ref{fig:rawhist} shows the distribution of raw scores, separated by Seal verdict, for the scored production population using the frozen production constants. The failing population is concentrated at the lower end of the scale but remains continuously ordered. Its median raw score is $8.9 \times 10^{-3}$, only 0.002\% of rows have a raw score $\geq 0.99$, and just 18 Seal-failing rows reach $\geq 0.90$. All of these are display-capped below 80 by the mapping defined in \S\ref{sec:display}, and the verdict-consistency invariant holds without a single violation across all 359{,}062 rows.

A total of 3{,}848 rows (1.07\%) pass all five validation gates. These are concentrated in long-history daily-bar runs, where sufficient statistical evidence can be accumulated, and their raw scores range from 0.39 to 0.9998, compared with a failing 99th percentile of 0.64.

Panel (b) shows only the 516 \emph{winners}---the highest-scoring result from each optimization run, corresponding to what a user actually sees. These scores are shifted substantially upward relative to the full production population, indicating that the strong compression near the bottom of the raw-score scale is mainly a property of the complete candidate population rather than of the strategies ultimately presented to users.

The figure is intended only to demonstrate that the construction avoids score saturation while preserving ordering among failing strategies. It should not be interpreted as describing the typical quality of production strategies.

\paragraph{Stability and weighting.}
The scored ledger is concentrated in a small number of high-volume operators,
with the top five accounting for more than 99\% of all rows. As a result,
the pooled row-weighted estimator and a run-weighted alternative (where each
search contributes equally) produce different estimates. The run-weighted
dispersion of the corrected DSR margin is 0.63, compared with 1.128 for the
pooled estimator.

We retain the pooled row-weighted value because it follows the convention
used in previous vintages and maintains comparability across the platform.
The run-weighted result is included in the artifact as a diagnostic measure,
allowing the effect of operator concentration to be quantified and monitored.

Constants of this type are defined relative to the underlying population.
Therefore, absolute scores are comparable only within the same calibration
vintage. Any recalibration is treated as a versioned change: the constants
are tied to a specific commit, the derivation script and frozen artifact are
stored in the repository, and a regression test detects any unintended drift.

\section{Predictive Validation in a Synthetic Ground-Truth
Environment}\label{sec:synthetic}

The key question for a robustness score is whether it ranks out-of-sample survival more effectively than existing alternatives. Because the production ledger records the validation gates but not the future performance of individual strategies, we address this question in a controlled synthetic environment with known ground truth, following the methodology of \citet{arian2024backtest}.

\subsection{Design}\label{sec:synth-design}

We generate $n = 2000$ synthetic strategies. Each strategy is the winner of a simulated parameter search that evaluates $n_{\mathrm{trials}} \sim \mathrm{LogUniform}(20, 5000)$ candidates over $S = 10$ CPCV folds of a $T$-bar daily history. Half of the searches contain a single genuine candidate with a true annualized Sharpe $\sim U(0.5, 2.5)$, while the remaining searches contain only noise. In every case, the in-sample best candidate is selected, reproducing the selection bias that the validation gates are designed to correct.

The complete validation battery is then applied to the selected winner using the production implementation: DSR with the $(\kappa+2)/4$ variance term and the persisted pre-$\Phi$ statistic, true CSCV PBO computed from the candidate-by-fold matrix, circular-block-bootstrap SPA, MinTRL, and the regime composite. These are combined into the MinervaScore raw score $r$ using the frozen calibration constants described in \S\ref{sec:calibration}. Throughout this section, the scorer labeled ``MinervaScore'' refers to this continuous raw score, following the naming convention introduced in \S\ref{sec:display}. Ground truth is available in two forms: a binary label indicating whether the selected winner is the genuine candidate, and a realized out-of-sample Sharpe ratio obtained from a fresh $T$-bar draw at the winner's true mean. The baselines are DSR alone, the number of validation gates passed, and a return-series proxy for the GT-Score \citep{sheppert2026gtscore}.

It is important to clarify what this comparison measures. All methods are evaluated on the same task: ranking selected strategies according to whether they contain genuine edge. This is the task of a post-search validator, not of a search objective. The GT-Score was designed to guide the optimization process (\S\ref{sec:pipeline}), rather than to validate the final strategy. Our comparison therefore evaluates how well a search objective performs when it is used as a post hoc quality score. We include it because it is a familiar curve-quality score and a natural
continuous benchmark for practitioners. The results should therefore not be interpreted as showing that the MinervaScore is a better search objective. Instead, they show that, when used as a validation score, it remains
competitive with the tested alternatives and more stable than
the other composite scores as the search problem becomes more difficult.

One simulation parameter deserves explicit mention. The history length $T$ is fixed at $1260$ bars (five years of daily data), because this setting reproduces the distributional property required for the study: a genuine-edge prevalence of $\approx 0.31$. When a genuine candidate is present, it wins roughly three-fifths of the searches, while the remaining searches are won by lucky noise. An earlier internal version of this experiment was conducted under a configuration that was not fully preserved, so absolute performance values cannot be compared across versions. The results reported here are therefore based entirely on the regenerated experiment.

\subsection{Results}\label{sec:synth-headline}

\begin{table}[t]
\centering
\small
\begin{tabular}{lccc}
\toprule
Scorer & \shortstack{Genuine edge\\(AUROC)} & \shortstack{OOS survival\\(AUROC, $> 0.5$)} & \shortstack{Spearman\\(OOS SR)}\\
\midrule
MinervaScore          & 0.9890 & 0.863 & 0.611\\
DSR alone             & 0.9880 & 0.862 & 0.609\\
GT-Score (proxy)      & 0.9860 & 0.862 & 0.608\\
Gates passed (count)  & 0.9597 & 0.850 & 0.592\\
\bottomrule
\end{tabular}
\caption{Predictive performance at the headline difficulty ($n=2000$,
$T=1260$; base rate of genuine edge 0.306), under the corrected DSR null
benchmark and the frozen constants of \S\ref{sec:calibration}.
Paired-bootstrap AUROC gaps (MinervaScore $-$ baseline, $B = 1000$):
$+0.0292$ $[+0.0231, +0.0360]$ against gates-passed and $+0.0030$
$[+0.0011, +0.0056]$ against the GT-Score proxy. The gap against DSR
alone is $+0.0010$ $[+0.0001, +0.0020]$: the interval is nominally
positive, but the effect is a tenth of a percentage point of AUROC and
$P(\text{gap} > 0) = 0.98$ against $\ge 0.999$ for the other two, so we
read it as a tie at this difficulty rather than as a lead.}
\label{tab:synthetic}
\end{table}

The comparison in Table~\ref{tab:synthetic} changes under the corrected null benchmark. The MinervaScore scores slightly above the GT-Score proxy, with a positive
bootstrap interval of $+0.003$ $[+0.001, +0.006]$, and shows a larger improvement
over the gates-passed baseline. The closest competitor is the corrected DSR-alone baseline, which achieves an AUROC of 0.988, within one bootstrap standard error of the full composite. This is expected, since correcting the deflation benchmark makes the DSR the main source of discrimination in the validation battery. Because the two scores are so close, a single comparison is not sufficient to distinguish them. The difficulty sweep presented below provides a more informative comparison.

\subsection{Difficulty sweep: worst-case regret}
\label{sec:synth-sweep}

\begin{figure}[t]
\centering
\includegraphics[width=0.9\linewidth]{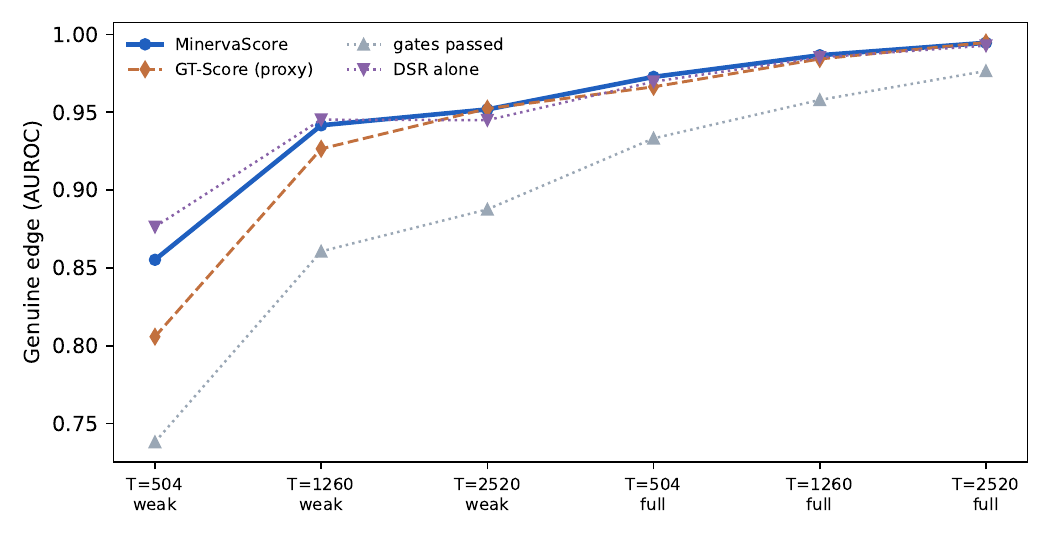}
\caption{AUROC against the genuine-edge label across the pre-specified
difficulty grid ($n = 1000$ per cell, all cells shown).}
\label{fig:sweep}
\end{figure}

We repeated the study over a pre-specified $3 \times 2$ grid, combining history lengths $T \in \{504, 1260, 2520\}$ with genuine-edge strengths of $\mathrm{SR} \sim U(0.25, 1.25)$ (weak) and $U(0.5, 2.5)$ (full). Each of the six cells contains $n = 1000$ simulated strategies, and all results are reported (Fig.~\ref{fig:sweep}).

No scorer performs best in every setting. The GT-Score proxy loses much of its discrimination when histories are short and true edges are weak, dropping from 0.995 in the easiest cell to 0.806 in the hardest. The gates-passed baseline performs worse than the other methods in every cell.

Among the composite scores, the MinervaScore has the smallest worst-case AUROC regret, \textbf{0.021}, compared with 0.071 for the GT-Score proxy and 0.138 for gates passed. The corrected DSR alone is also robust across difficulty levels. Its worst-case regret is only 0.009, the smallest overall, and it achieves the best performance in the hardest cell. Once the null benchmark no longer over-deflates long histories, the DSR remains reliable across the entire grid. The advantage of the MinervaScore is therefore not a larger worst-case discrimination margin but its broader construction (\S\ref{sec:synth-headline}, Table~\ref{tab:compare}), which combines the strongest validation component with the additional coverage of the full validation battery and the verdict guarantee.

In the hardest cell (base rate 0.03), none of the methods predicts the realized out-of-sample Sharpe ratio (Spearman $\approx 0$ for all methods), and we report this null result rather than excluding it.

\subsection{How the score and its components respond to search size and
history length}\label{sec:synth-behavior}

\begin{figure}[t]
\centering
\includegraphics[width=\linewidth]{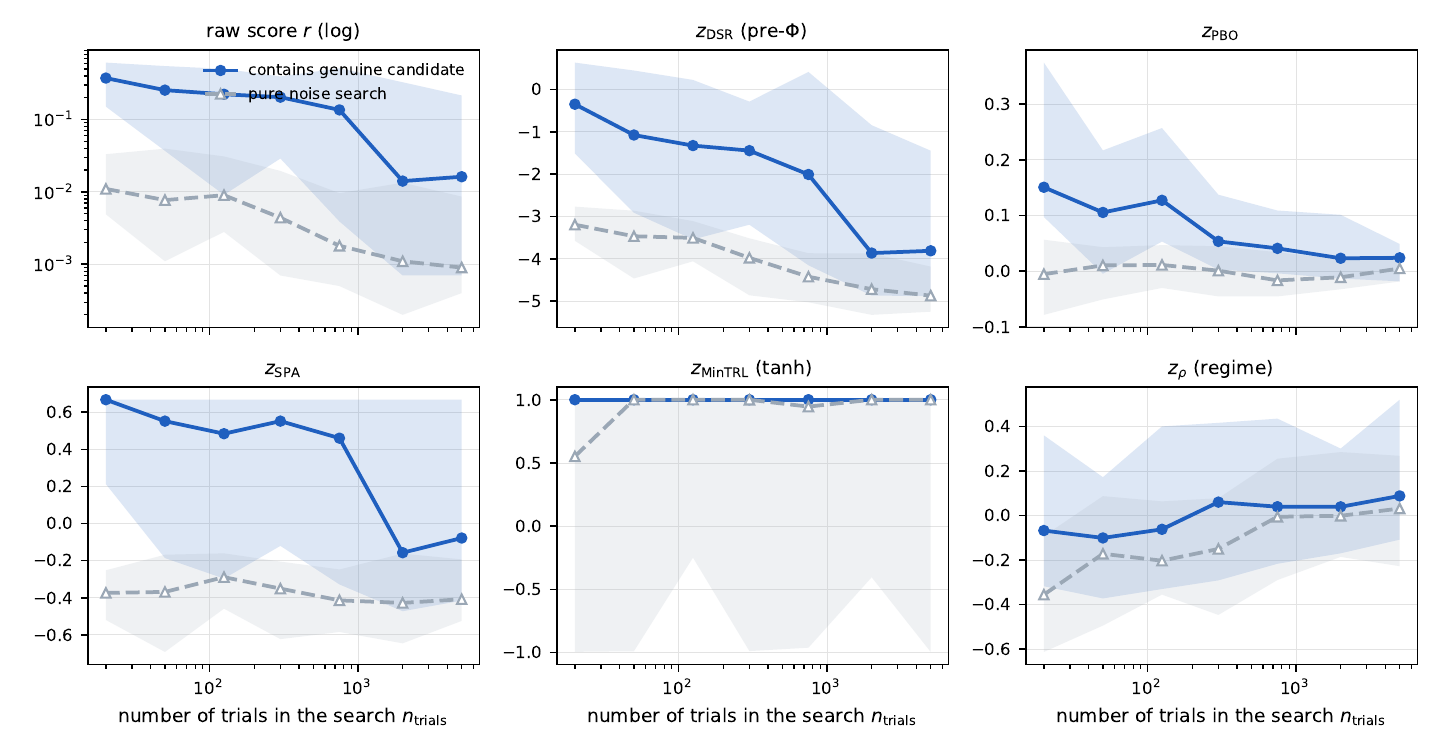}
\caption{Raw score and per-gate margins versus the number of trials in the
search ($T = 1260$ fixed; median and IQR over 60 searches per population per
cell). Score medians below $10^{-4}$ are plotted at the axis floor.}
\label{fig:behavior-ntrials}
\end{figure}

\begin{figure}[t]
\centering
\includegraphics[width=\linewidth]{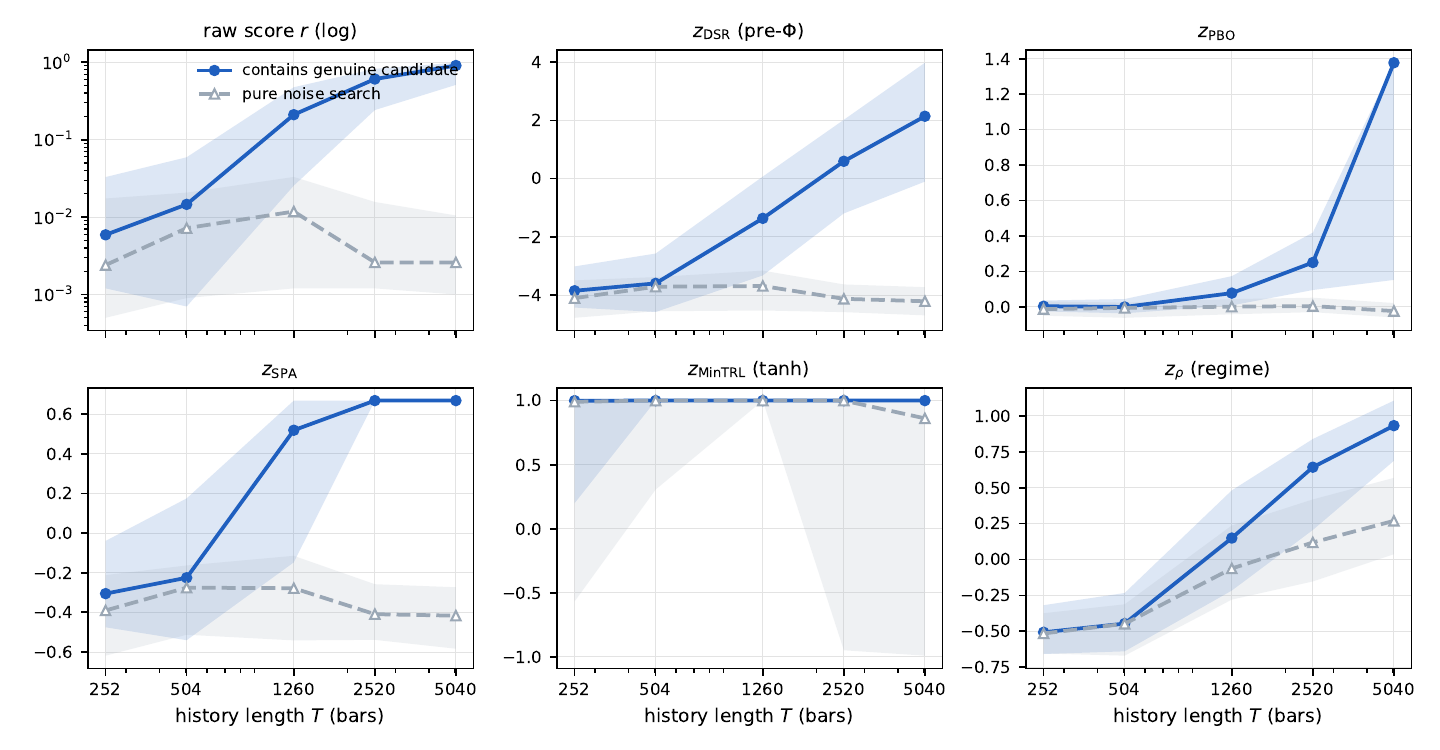}
\caption{Raw score and per-gate margins versus history length
($n_{\mathrm{trials}} = 300$ fixed; median and IQR over 60 searches per
population per cell). Score medians below $10^{-4}$ are plotted at the
axis floor.}
\label{fig:behavior-tbars}
\end{figure}

The two parameters directly controlled by the user are the search size and the length of the backtest window. Figures~\ref{fig:behavior-ntrials} and~\ref{fig:behavior-tbars} examine these two factors separately by varying one while keeping the rest of the simulation environment fixed. For each setting, we report the median and interquartile range of the raw score and each standardized gate margin, distinguishing between searches that contain a genuine candidate and searches that contain only noise (60 of each per setting). All results are computed using the production gate functions.

\textbf{Search size (Fig.~\ref{fig:behavior-ntrials}).} Larger searches receive lower scores, as intended. Across all search sizes, the median raw score of pure-noise searches remains at or below 0.011, while the DSR margin decreases by roughly $0.7\sigma$ for every tenfold increase in the number of trials, reflecting the increasing penalty for search effort. The median score of searches containing a genuine candidate also declines, from 0.37 at 20 trials to 0.016 at 5000 trials. This reduction is appropriate rather than overly conservative because, as the search becomes larger, the genuine candidate is more likely to be outperformed by lucky noise, and the median realized Sharpe ratio of the selected winner decreases from 1.4 to 0.5.

Accordingly, a winner selected from a 5000-trial search deserves less confidence than one selected from a much smaller search, and the MinervaScore reflects this. The Seal rate for searches containing a genuine candidate falls from approximately 20\% at small search sizes to 7\% when $n_{\mathrm{trials}} \ge 2000$, whereas no strategy from the pure-noise population achieves the Seal at any search size. The SPA margin exhibits the same pattern of dilution. By contrast, PBO and the regime composite remain almost unchanged as the search size increases, which is expected because both are evaluated from the consistency of the selected winner itself rather than from the size of the search that produced it.

\textbf{History length (Fig.~\ref{fig:behavior-tbars}).} Longer histories improve discrimination, and under the corrected null benchmark the two populations behave as expected. The DSR margin for the pure-noise population remains essentially constant as $T$ increases (median $-3.7$ to $-4.2$ over a twenty-fold increase in history length). This is the behavior expected under a correctly specified null model: additional history does not make a noise strategy more or less convincing. The flat trend also illustrates the effect of the benchmark correction. Under the previous fixed benchmark, the same margin decreased steadily to $-7.5$, an artifact of treating every backtest as if it covered only one year.

The genuine population behaves very differently. Its median DSR margin increases from $-3.9$ at $T = 252$ to $+2.1$ at $T = 5040$ as statistical evidence accumulates. Over the same range, the median raw score rises from 0.006 to 0.91, and the Seal rate increases from zero (one year of data, where the evidence is correctly insufficient) to 17\% after five years and 73\% after twenty years. By contrast, no strategy in the pure-noise population achieves the Seal at any history length.

The PBO and regime margins show the same qualitative pattern, increasing for genuine strategies ($+1.4$ and $+0.9$, respectively, at $T = 5040$) while remaining essentially unchanged for noise. Under the corrected benchmark, additional history therefore strengthens the evidence for genuine strategies without affecting noise strategies. The MinTRL margin remains at its tanh bound throughout both parameter sweeps and becomes informative only for very short histories or datasets with very few observations.

\subsection{Calibration of the score as a probability}
\label{sec:synth-calibration}

\begin{figure}[t]
\centering
\includegraphics[width=0.6\linewidth]{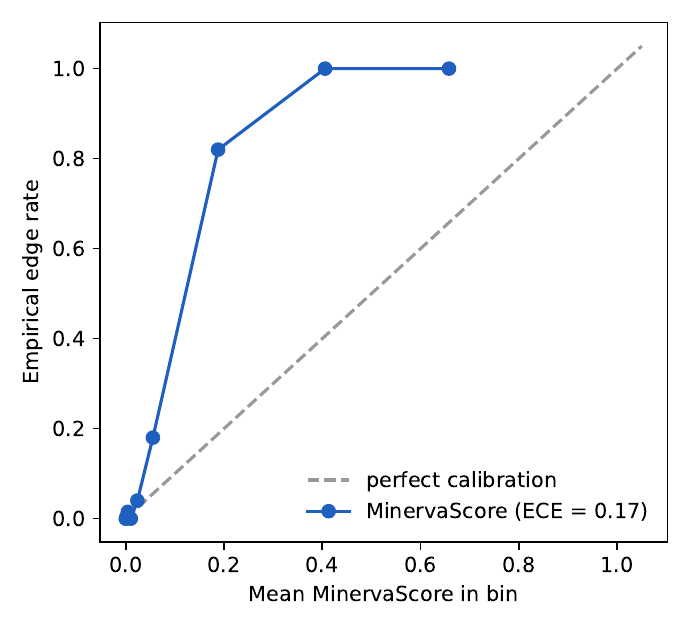}
\caption{Reliability diagram against the genuine-edge label at the headline
cell (equal-count bins). The curve is monotone, so the score ranks, but it
sits far from the diagonal (ECE 0.174).}
\label{fig:reliability}
\end{figure}

The MinervaScore is mechanically similar to a Platt-style transformation \citep{platt1999probabilistic}, but without the calibration step. The offset $c$ plays the role of Platt's bias term, but it is fixed rather than fitted to observed outcomes. As a result, $\Phi(S-c)$ should not be interpreted as the probability that a strategy has genuine edge. Figure~\ref{fig:reliability} confirms this: at the headline simulation setting, the reliability diagram yields an expected calibration error of 0.174. The MinervaScore is therefore intended as a ranking measure rather than a calibrated probability. Achieving proper probabilistic calibration in the sense of proper scoring rules \citep{gneiting2007scoring} would require the forward-outcome data discussed in \S\ref{sec:conclusion}.

\subsection{Using the environment to test proposed changes to the score}
\label{sec:synth-srvariance}

The ground-truth environment is also used to evaluate whether a proposed modification to the score improves the construction. \citet{bailey2014dsr} recommend estimating $\mathrm{Var}[\{\widehat{SR}_n\}]$ directly from the set of trials. When no trial panel is available, our implementation historically used a fixed fallback of $\mathrm{Var}[\widehat{SR}] = 1$ in annualized units. This was a house convention rather than a prescription of \citet{bailey2014dsr}; their use of $V=1$ appears only in illustrative examples.

We evaluated two alternatives to this fallback. The first was an empirical per-family variance estimated from the sibling candidates produced by each search, motivated by the observation that real strategy families may have lower dispersion than assumed by the fixed constant. In a labeled environment containing both genuine and spurious families, however, this modification \emph{reduced} the separation between genuine and spurious strategies (AUROC $0.96 \to 0.50$), with the result remaining stable across the parameter sweep. The reason is that the dispersion of sibling Sharpe ratios is dominated by estimation noise rather than by information related to the ground-truth label. Normalizing by this quantity therefore introduces additional noise into an otherwise monotonic statistic. We consequently rejected this modification as score inflation.

The second modification replaced the fixed constant with the null sampling variance of the annualized Sharpe estimator from \citet{lo2002sharpe}, namely $1/\mathrm{years}$ (equivalently $1/T$ per bar). The original fixed value of $1.0$ corresponds to this null variance only for a one-year backtest and therefore overstates the noise-only expected maximum for longer histories. Unlike the per-family estimate, this correction is formula-based and does not introduce estimation noise. Evaluated in the same ground-truth environment, it \emph{improved} separation (MinervaScore AUROC $0.9855 \to 0.9894$; DSR-alone $0.9735 \to 0.9880$). At the same time, the DSR gate's false-positive rate on pure-noise strategies remained exactly zero, while its true-positive rate on genuine strategies increased from $1\%$ to $41\%$.

The dispersion constants available at the time of the experiment were used for this A/B comparison. Therefore, the frozen-constant headline reported in Table~\ref{tab:synthetic} ($0.9890$) and the corrected composite differ in the fourth decimal place. The single-gate baselines match exactly because they do not depend on the dispersion constants. We therefore adopted the null-benchmark fallback, while older code paths with unresolved units continue to use the historical constant.

This comparison also defines the criterion used for methodological changes: a modification that increases scores must first demonstrate improved separation with respect to ground truth, rather than simply producing a higher pass rate.

\subsection{Limitations of the synthetic evidence}
\label{sec:synth-limits}
(i) The absolute performance levels should be viewed as optimistic because the synthetic environment more closely satisfies the validation gates' assumptions than actual financial markets. The relative ranking of the approaches and their regret profiles are the outcomes that are anticipated to generalize.

(ii) Under the updated null benchmark, the binary Seal is no longer
degenerate. In this simulation, every Seal pass occurs on a strategy with
genuine edge, and the admission rate is 8.15\%. This supports the
interpretation of the pass/fail criterion within the synthetic environment.
However, this admission rate should be considered an upper bound rather
than a production forecast because the simulated environment is still
cleaner than actual markets.

(iii) Using a zero benchmark, the GT-Score baseline is implemented as a return-series proxy of the published trade-level objective. The comparisons in the synthetic study are unaffected by any of these restrictions. However, they are one of the primary motivations for carrying out the independent real-market assessment described in \S\ref{sec:realmarket}.

\section{A Pre-Registered Real-Market Study}\label{sec:realmarket}

\subsection{Design}\label{sec:rm-design}
A null was obtained from an earlier real-market check of the prior score design, but this null proved to be unhelpful. Because market beta dominated the bull-regime outcome and the score had virtually no cross-sectional range to predict with (scores confined to $\approx 0.01$--$0.04$; per-ticker inter-quartile range $\approx 0.001$), neither the score nor any baseline could predict out-of-sample Sharpe on two index universes. The study design itself contributed to some of that range restriction. The per-ticker rows of a universe-mode optimization are the same strategy re-evaluated on different names, inheriting the validation values of the parent, because it does a single parameter search for the entire index. Regardless of the score's advantages, a strategy that generates nearly constant predictors cannot demonstrate predictive power.

Both the instrument and the design are modified in this study. The unit of analysis is a unique searched strategy, and the deflation inputs (trial counts, candidate-by-fold matrices) describe the search that actually produced each strategy. The primary analysis (\S\ref{sec:rm-primary}) optimizes each ticker independently---one search, one winner, and one gate battery per name. 2016-04-19 to 2018-04-19 (5-minute bars) is the in-sample timeframe. The out-of-sample window, 2018-04-20 to 2020-04-20, is \emph{sealed}: a series of fixed-parameter backtests of the pre-registered winners opened it exactly once, and no experiment of any type had ever been conducted on it. Prior to unsealing, enrollment, exclusion rules, the COVID handling rule, and the analysis plan were frozen; the calibration constants are pinned at the commit that froze them. The population spans two liquidity tiers of the same market because tickers are drawn from two index universes: the S\&P 500 and the S\&P MidCap 400 (henceforth referred to as S\&P 400), which are the standard U.S. equity benchmarks covering, respectively, roughly the 500 largest listed U.S. companies by market capitalization and the mid-capitalization tier below them.
Power was explicitly budgeted: $n = 400$ strategies (the first 100 elements
of each index in alphabetical order, two model families) provide about 80\%
power at two-sided $\alpha = 0.05$ under a Fisher-$z$ approximation to
detect a score--outcome rank correlation of $\rho \approx 0.14$, which we
consider to be the smallest practically interesting effect; the same
calculation yields $\rho \approx 0.15$ at the 352 strategies that survived
enrollment. Every analysis in this part uses the exported trial ledgers
from the Minerva platform, and all in-sample searches, the universe-mode
cells of \S\ref{sec:rm-universe}, and the sealed-window confirmatory batch
were submitted and carried out as standard jobs.

\subsection{Primary result}\label{sec:rm-primary}

The pre-registered analysis, with both the enrolled strategies and their in-sample predictor values fixed before the sealed evaluation window was opened, yields a null result. Across the 352 enrolled strategies, the pooled Spearman correlation between the in-sample raw score $r$ (no enrolled strategy achieved the Seal, so the displayed MinervaScore produces the same ordering) and the realized out-of-sample Sharpe ratio is $\rho_s = 0.013$ (95\% bootstrap CI $[-0.094,\, 0.121]$; one-sided permutation $p = 0.40$; AUROC for $\mathrm{Sharpe}_{OOS}>0$ equal to 0.496). The pre-registered decision criterion ($p < 0.05$ \emph{and} a confidence interval with lower bound $>0$) is therefore not satisfied, and we make no claim of real-market predictive power. The pre-registered DSR-only baseline is similarly uninformative
($\rho_s = 0.006$), indicating that none of the tested predictors shows
evidence of a relationship with out-of-sample outcomes on this dataset.

This null result differs from the earlier study in two important ways. First, it is informative. The previous analysis suffered from an almost complete lack of variation in the predictor (per-ticker IQR $\approx 0.001$), making meaningful discrimination impossible. In contrast, the predictor in the present study has substantial variability (IQR $0.057$), and the sample size provides the planned statistical power. Consequently, $\rho_s \approx 0.01$ reflects the absence of an association rather than a restriction of range. The two-scale transformation made this analysis possible, and the result is still null.

Second, the out-of-sample outcomes themselves provide little signal. The median realized out-of-sample Sharpe ratio is $-0.15$, and only 45\% of strategies achieve a positive out-of-sample Sharpe. The study therefore evaluates a population with very little surviving edge---mean-reversion strategies trained on 2016--2018 data and evaluated over the 2018--2020 period, including the COVID market crash. This finding is consistent with the synthetic experiments rather than contradictory to them. Section~\ref{sec:synthetic} shows that the MinervaScore distinguishes true signal from lucky outcomes when genuine edge exists and is detectable. In this real-market population, little genuine edge appears to remain, and accordingly the score does not produce a signal where none exists.

The null result should also be interpreted with appropriate caution. It does not invalidate the construction itself, since the point-mass correction, the calibration procedure, the verdict-consistency invariant, and the synthetic validation are each supported independently. Nor does it imply that the score is without value. A conservative validation metric should avoid certifying strategies when genuine edge is largely absent, and its continuous score should not create artificial rankings within a population dominated by noise. The main conclusion is therefore more limited: while the MinervaScore successfully ranks genuine edge in a controlled environment with known ground truth, this real-market study provides no evidence of forward predictive power, and we make no stronger claim.

One of the two model families shows a nominally positive within-family correlation (Bollinger: $\rho_s = 0.134$, one-sided $p = 0.036$), whereas the other does not (Kalman: $\rho_s = -0.003$). Because only two model families were examined, the confidence interval includes zero, and no correction for multiple comparisons was applied, this result does not satisfy the pre-registered confirmatory criterion. We report it for completeness but draw no conclusion from it. Selecting the favorable model after observing the results would itself constitute the type of outcome-dependent selection that the validation framework is designed to prevent.

The pre-registered analysis excluding the COVID crash rules out the most obvious confounding explanation. Recomputing each strategy's realized Sharpe ratio using only the pre-crash portion of the out-of-sample window ($\leq$ 2020-02-19) reproduces the production full-window Sharpe with a median absolute error of $4\times10^{-4}$. The pooled correlation remains unchanged at $\rho_s = 0.013$, while the median truncated out-of-sample Sharpe is still $-0.15$, with 45\% of strategies remaining positive. The null result is therefore not caused by the COVID crash. These strategies were already unprofitable before February 2020, and the absence of a relationship between the score and subsequent performance does not depend on that market regime. The within-family results are likewise unchanged (Bollinger: $0.135$; Kalman: $-0.003$).

\paragraph{Exploratory only: the null survives the corrected instrument.}
After completing this study, we corrected the null benchmark used by the DSR gate (\S\ref{sec:synth-srvariance}). Because the pre-registered predictor had already been computed with the previous benchmark, the confirmatory analysis reported above remains unchanged by design. As an \emph{exploratory} analysis only---not a confirmatory one, since re-scoring after observing the outcomes is precisely the type of post-hoc analysis that pre-registration is intended to prevent---we re-scored 340 of the 352 locked strategies using the corrected DSR and repeated the pooled analysis on the same subset.

The result remains null: $\rho_s = +0.011$ (one-sided permutation $p = 0.42$, AUROC 0.503), compared with $+0.003$ for the original locked predictor on the identical subset. The corrected score changes the ranking very little, with an old-versus-new predictor rank correlation of 0.983. This is expected because the enrolled strategies were generated under nearly identical conditions (single configurations, identical backtest windows, and identical bar intervals), leaving little variation for the corrected null benchmark to affect.

Rather than weakening the conclusions, this result reinforces the argument made in \S\ref{sec:conclusion}: evaluating the corrected score on a more heterogeneous population, spanning diverse search configurations and containing a meaningful amount of genuine edge, is necessary to assess its real-market predictive value.

\paragraph{What these results mean in practice.}
In practical terms, the results imply three points. First, the Seal should
be read as a certification criterion rather than as a trading rule. For a
single-name strategy chosen from thousands of candidates based on two years
of data, certification is expected to be rare: the history-length sweep of
\S\ref{sec:synth-behavior} shows that the evidence requirement can exceed
what such a history contains even when the edge is real. A low pass rate on an edge-poor population is therefore expected. Second, the continuous score is the part of the framework evaluated as a ranking in
\S\ref{sec:synthetic}: relative standing can still convey information when
the binary verdict alone is too coarse.
Third, this article does not establish a decision rule of the form ``trade
above score $X$.'' Establishing such a threshold would require forward
outcomes on a population with genuine edge, which the present real-market
population lacked (\S\ref{sec:conclusion}).

\subsection{Secondary analysis: universe-mode optimization}
\label{sec:rm-universe}

Universe-mode optimization addresses a different but complementary question: can a \emph{single} parameter configuration generalize out-of-sample across an entire index? This is a substantially stronger notion of robustness than achieving edge on individual securities, since the same parameter set must produce a non-negative out-of-sample Sharpe ratio simultaneously across hundreds of stocks. We therefore pre-registered one universe-mode optimization for each model $\times$ universe combination, using the same time window, bar interval, and optimizer settings as in the primary study, and report the results descriptively.

\begin{table}[t]
\centering
\small
\begin{tabular}{llrlc}
\toprule
Cell & Names & Surviving candidate rows & Headline candidate & Seal \\
\midrule
Kalman $\times$ S\&P 500    & 472 & 2{,}920 & SR 2.10, 13 trades, score 0.102 & fail \\
Bollinger $\times$ S\&P 500 & 472 & 54      & SR 1.21, 9 trades, score 0.040 & fail \\
Kalman $\times$ S\&P 400    & 345 & 1{,}090 & SR 2.01, 9 trades, score 0.063 & fail \\
Bollinger $\times$ S\&P 400 & 345 & 27      & SR 2.12, 15 trades, score 0.153 & fail \\
\bottomrule
\end{tabular}
\caption{Universe-mode (shared-parameter) cells, all executed on the same
platform version. ``Names'' is the effective universe recorded by the run.
Candidate scores across the four cells span up to 0.38; no cell produces a
Seal pass, and every headline candidate trades 9--15 times in two years ---
too little evidence for statistical certification.}
\label{tab:universe-cells}
\end{table}

Three observations emerge from Table~\ref{tab:universe-cells}. First, parameter configurations that generalize across an entire index do exist, but the supporting evidence is consistently weak. Every cell produces at least one surviving configuration, yet the best candidate executes only 9--15 trades over two years across the whole index. None therefore satisfies the evidence requirements for the Seal. This contrasts with the per-ticker optimizations in the primary study, which routinely produce candidates with hundreds of trades and complete validation information using the same time window and bar interval. At this frequency, per-stock edge is relatively common, whereas a single configuration that works across an entire index is much rarer, and the available evidence is insufficient to evaluate it statistically. Accordingly, these cases are reported as ``insufficient evidence'' rather than as either passing or failing the validation criteria.

Second, this pattern is consistent across both market universes and both model families. The Kalman models produce 20--50$\times$ more evidence-passing per-ticker results than the Bollinger models on both indices, a difference already documented in the primary per-ticker analysis.

Third, the two-scale transformation restores useful score variation even in this deliberately challenging setting. Candidate raw scores reach values as high as 0.38, whereas the previous version of the score compressed comparable populations into the narrow range of 0.01--0.04. Because the secondary study includes only four cells, it is descriptive rather than inferential and does not support correlation analyses. It nevertheless answers an important practical question: why not optimize a single parameter configuration for an entire index? The reason is that the resulting strategies generate too little evidence to support statistical certification.

\paragraph{Provenance.}
The platform's handling of index constituents with incomplete historical coverage changed during the course of this work. Current constituent lists include securities that were listed or changed identity within the evaluation window, and the platform now evaluates each constituent only over the period for which data are available, recording the effective coverage span for every result.

The initial universe-mode experiments on the S\&P 400, performed before
this platform update, produced no surviving candidates. After the coverage
logic was revised, the experiments were repeated using the updated platform
and yielded the results reported in Table~\ref{tab:universe-cells}. The
original zero-survivor result was therefore not reproduced and is
superseded by the updated results, although both are reported for
completeness. All four cells in Table~\ref{tab:universe-cells} were
generated using the same platform version.

For the per-ticker analyses extracted from the universe-mode runs, we apply the same pre-registered coverage criterion as in the primary study: securities with less than 90\% coverage of the evaluation window are excluded (for example, VICI, listed in 2018-02, with approximately 10\% coverage), and the effective universe is recorded for every run.

\section{Robustness Audit}\label{sec:audit}

A score intended to measure robustness should be tested accordingly. We
report five checks, all computed under the frozen constants on the scored
production population of 359{,}062 statistic-bearing trial-ledger rows
across 536 optimization runs. Three checks provide empirical evidence
(E3--E5), while two verify implementation properties (E1--E2).

\subsection{E1: no saturation}

A robustness grade becomes uninformative if a large fraction of the
population accumulates at the top of the scale. This does not occur here:
on the production population, the fraction of raw scores $\ge 0.99$ is
\textbf{0.002\%}, the failing median is $8.9 \times 10^{-3}$, and only
\textbf{18} Seal-failing records reach a raw score $\ge 0.90$. Each of
these records is display-capped below 80 by the band projection of
\S\ref{sec:display}, so no failing record can be displayed as passing. We
report the raw-scale tail count because the raw score is persisted and an
operator inspecting it should be aware that this tail exists. For
comparison, the in-house predecessor replaced by this construction, which
used raw gate probabilities without margin encoding, offset correction, or
Seal gating, assigned roughly one third of a comparable population a score
of $\ge 0.99$.

\subsection{E2: verdict-consistency}

The invariant \eqref{eq:invariant} holds with \textbf{zero violations}
across all 359{,}062 records, and was re-verified at the certification
threshold of 80 after the percentile-based display revision. No
Seal-failing record displays at or above 80 (the failing maximum is 79),
and no sealed record displays below 80 (sealed range 82--100, median 91).
This check acts as a unit test for truncation or floating-point errors in a
property that is enforced by construction; it provides engineering
assurance rather than evidence about the strategies themselves.

\subsection{E3: sensitivity to the weights and the offset}

\begin{table}[t]
\centering
\small
\begin{tabular}{lcc}
\toprule
Perturbation ($\pm 20\%$) & Spearman $\rho$ vs.\ baseline & $\max |\Delta r|$\\
\midrule
$w_{\dsr}$    & 0.9974 & 0.099\\
$w_{\pbo}$    & 0.9991 & 0.033\\
$w_{\spa}$    & 0.9995 & 0.074\\
$w_{\mintrl}$ & 0.9999 & 0.012\\
$w_{\rho}$    & 0.9992 & 0.041\\
offset $c$    & 1.0000 & 0.040\\
\bottomrule
\end{tabular}
\caption{Rank stability under $\pm 20\%$ perturbation of each weight
(renormalized over the active mask) and of the offset $c$, over the
359{,}062-row production population. The offset is rank-invariant exactly,
as required analytically.}
\label{tab:sensitivity}
\end{table}

Rankings remain stable under $\pm 20\%$ perturbations of all weights
(Table~\ref{tab:sensitivity}), while the offset $c$ leaves the ranking
exactly unchanged, as shown in \S\ref{sec:offset}. Among the six
parameters, the DSR weight is the most sensitive: its worst-case Spearman
correlation is 0.9974, compared with $\ge 0.9991$ for every other
perturbation. This sensitivity is expected because the DSR margin is
computed on the pre-$\Phi$ scale, which spans a wider range than the other
margins. Even in this worst case, the induced ordering remains essentially
unchanged, and no perturbation within $\pm 20\%$ produces a meaningful
re-ranking of the population.

\subsection{E4: the sealed set, and boundary exposure}

On the production population, \textbf{3{,}848} rows (1.07\%) clear all five
gates. These rows span a raw-score range of 0.391--0.9998, compared with a
failing p99 of 0.64. They are not uniformly distributed across the ledger:
they are concentrated in six long-history daily-bar optimization runs,
where the evidence requirement can be met
(\S\ref{sec:synth-behavior}). Boundary exposure is limited: 18.8\% of
records lie within $0.05\,\sigma_k$ of at least one gate threshold, and
tightening every threshold by $0.05\,\sigma_k$ against the strategy changes
the verdict of 311 of the 3{,}848 sealed rows (8.1\%). Thus, most sealed
strategies retain a margin above the thresholds. Nevertheless, strategies
close to a boundary should be evaluated using the underlying raw score
$r$, which is persisted alongside the displayed band.

\subsection{E5: realized per-gate contribution versus the weights}

\begin{table}[t]
\centering
\small
\begin{tabular}{lccc}
\toprule
Gate & Weight & Share at the perfect point & Share over the population\\
\midrule
DSR    & 0.35 & 32.7\% & 76.9\%\\
PBO    & 0.25 & 11.7\% & 4.6\%\\
SPA    & 0.20 & 13.1\% & 8.1\%\\
MinTRL & 0.10 & 3.4\%  & 2.8\%\\
Regime & 0.10 & 39.2\% & 7.6\%\\
\bottomrule
\end{tabular}
\caption{Per-gate share of the aggregate: at the theoretical perfect point
(numerator decomposition of $w^{\top}z^{\star}$) and as the mean absolute
contribution over the 359{,}062-row production population.}
\label{tab:contribution}
\end{table}

Each gate's real contribution is measured in Table~\ref{tab:contribution}, which differs from what its nominal weight indicates \citep{paruolo2013ratings}.
The majority of the ordering effort is done by the DSR, which completely dominates the blend over the population (76.9\% of mean absolute contribution on a 0.35 weight). This is because most strategies fail it by margins that the pre-$\Phi$ coordinate can represent. Due to its wide clamped-perfect logit margin, the regime composite contributes 39.2\% on a nominal weight of 0.10, whereas the DSR provides 32.7\% at the theoretical perfect point. There is still a discrepancy between nominal and effective weight. As \S\ref{sec:discussion} notes, the regime composite continues to be the most ad hoc component of the structure and has greater effect at the top of the scale than its weight suggests.

\section{Discussion}\label{sec:discussion}

\subsection{Where the layer fits, and where it must not be used}
Search, validation, and display serve different roles in the development
pipeline. A search objective, such as the GT-Score
\citep{sheppert2026gtscore} or the cross-validated Sharpe objective used in
the optimizer (\S\ref{sec:pipeline}), is designed to select a candidate.
The MinervaScore is different: it is computed only after selection, using
information that was not itself optimized. Feeding the MinervaScore back
into the search loop would make it another optimization target and would
expose it to the same regressional and extremal Goodhart failure modes
\citep{manheim2018goodhart} that the validation layer is meant to detect.
It should therefore be used for post-selection certification and reporting,
not as an in-loop prize.

\subsection{The adaptive-data-analysis hazard in a scoring service}

An optimizer is not necessary for a more subdued form of the same risk. The query sequence becomes adaptive and traditional held-out guarantees fail when a user submits a variant, reads its score, then selects the subsequent variant using the same data, as described by \citet{dwork2015reusable}. Because the gates are unable to view the variations a user attempted and
rejected, the workflow can then replicate the same selection-induced
overfitting problem the score identifies at the level of the analyst rather
than the optimizer. We therefore treat this as a first-order limitation of
the scoring workflow. The reusable-holdout discipline, query budgets, and a locked final test window that the iterative loop never touches are mitigations; the last of these is implemented for our own research in the sealed window of \S\ref{sec:realmarket}, and we refer to the remaining ones as necessary product work.

\subsection{Relation to the composite-indicator literature}

Two of the lessons learned from the mature methodology of collapsing several criteria into a single number are immediately applicable here \citep{saisana2005uncertainty}.
First, a robust composite should include an uncertainty-and-sensitivity analysis of its weights and aggregation, which is what \S\ref{sec:audit} does. This places the audit within standard composite-indicator practice. Second, the nominal-versus-effective-weight problem described by
\citet{paruolo2013ratings} also appears here. Table~\ref{tab:contribution}
shows that the realized contribution of each gate differs from its nominal
weight. The two-scale rescale and the corrected null benchmark reduce part
of this mismatch: for example, the DSR's share at the theoretical bound
rose from 6.4\% to 32.7\% across the two revisions. The mismatch is not
eliminated, however, since the regime composite still contributes 39.2\%
at the bound despite having a nominal weight of 0.10.

\subsection{The hard threshold and the dichotomy critique}
The post-$p$-value literature cautions against turning continuous evidence
into a hard threshold \citep{wasserstein2016asa, mcshane2019abandon}. This
concern applies here as well. In our setting, however, the Seal is used as
an explicit admissibility verdict, while the raw score $r$ remains available
for readers who prefer continuous evidence. Compared with the field's
tightening toward 0.05, our $\tau_{\spa} = 0.10$ is looser; a stricter
operator can set 0.05 without altering the structure.

\subsection{What the calibration fixes, and what it does not}
\label{sec:calibration-scope}

The construction distinguishes between quantities that define admissibility and quantities that describe the population used for calibration. The five gates and their thresholds $\tau_k$ are policy choices: they specify what is considered admissible evidence, are fixed in advance, and do not depend on the calibration population. By contrast, the dispersions $\sigma_k$, the correlation matrix $\Sigma_{\mathrm{eff}}$, and the quantile knots used for the percentile display are estimated from the population of strategies against which a result is evaluated.

This distinction means that recalibration does not change the pass/fail decision. The Seal is computed directly from the raw gate values and the thresholds $\tau_k$; no dispersion constant enters this step. Re-estimating $\sigma_k$ on a different population therefore leaves all Seal decisions unchanged. It can, however, change the ordering of strategies within a band and their percentile-based displayed scores. Recalibration changes the resolution of the continuous ranking, not the admissibility criterion.

The magnitude of this effect can be seen from the examples in \S\ref{sec:calibration}. Estimating $\sigma_{\pbo}$ on the clamp-free core rather than on the full, clamp-inclusive population changes it from $10.028$ to approximately $1.1$. Under the latter value, the fraction of rows with a score $\geq 0.98$ increases to $10\%$, instead of remaining a negligible tail. Similarly, weighting the calibration ledger by optimization run rather than by row changes the corrected DSR dispersion from $1.128$ to $0.63$. In both cases, the change comes from the reference population used to estimate the calibration quantities, not from a change in the gates themselves.

The constants reported in this paper therefore describe the operating population used for the reported results. An operator working with a different mix of strategies should expect different calibration values. The released derivation script can recompute these quantities from any trial ledger with the same schema, and the scoring function accepts $\sigma_k$, $\Sigma_{\mathrm{eff}}$, the weights, and the offset as arguments, allowing an alternative calibration to be supplied without changing the construction. We report one fixed calibration vintage so that results remain comparable across the platform; the reported values should not be interpreted as intrinsic properties of the method.

\subsection{Limitations}

\begin{enumerate}
\item \textbf{Predictive validation is strongest in synthetic ground truth;
the real-market result is an informative null.}
\S\ref{sec:synthetic} establishes discrimination and difficulty-robustness
in controlled environments; the pre-registered sealed-window study
(\S\ref{sec:realmarket}) returns a null result on a real population with
little detectable edge (pooled Spearman $0.013$, $p = 0.40$), so the study does not demonstrate forward predictive power there. The ECE
of 0.174 confirms that the score should be read as a ranking rather than as
a calibrated probability.

\item \textbf{The near-tie with the adjusted DSR is real.} The argument for the whole construction depends on the battery's coverage, the verdict-consistency invariant, and the audit; the corrected DSR alone matches the MinervaScore's discrimination at the headline difficulty and bears the smallest worst-case regret on the grid (Table~\ref{tab:compare}), not on consistent dominance.
\item \textbf{The choice is unaffected by the continuous machinery.} The five binary gates alone determine the displayed band; the continuous
machinery only orders strategies within a band.
\item \textbf{Scores are population-relative.} Recalibration is a versioned event, absolute scores are only comparable within a single vintage, and calibration constants are characteristics of a frozen calibration vintage.
\item \textbf{The sealed set is attainable but concentrated.} 
3{,}848 of 359{,}062 production rows (1.07\%) achieve the Seal, and these
rows are concentrated in six long-history daily-bar runs
(\S\ref{sec:audit}). The practical implication is that statistical evidence
accumulates with available history: under the corrected benchmark, the DSR
requirement at $N = 100$ trials corresponds to an annualized Sharpe of
$\approx 1.9$ on a five-year history and $\approx 1.3$ on a ten-year history,
compared with $\approx 3$ under the previous fixed constant. The real-market
populations of \S\ref{sec:realmarket} --- based on two-year windows and
largely lacking detectable edge --- produced no Seal passes, consistent with
this scaling. The sealed base rate is therefore a property of the searched
population, not an intrinsic property of the construction.

\item \textbf{The regime composite remains the most ad-hoc gate} (three
hand-set sub-weights, a fixed reference scale, and a logistic transformation)
and has a disproportionate influence near the upper end of the score range.
Its formulation remains an open area for improvement.

\item \textbf{Gate-input assumptions.} The DSR variance correction uses the
i.i.d.\ Mertens adjustment and does not account for serial correlation. The
deflation fallback variance now uses the Lo null sampling variance rather
than a fixed constant (family-level variance estimation remains rejected,
\S\ref{sec:synth-srvariance}), while call paths without resolvable history
units retain the legacy constant. The HAC Sharpe estimate is stored as an
audit field but is not included in the gate. The deflation count uses an
effective number of trials obtained through decorrelation of searched
parameters; the corresponding estimator remains an approximation.
\end{enumerate}

\section{Conclusion and Future Work}\label{sec:conclusion}

\subsection{What the score is for}
After running a backtest, a practitioner usually inspects drawdown, the
equity curve, and the Sharpe ratio. These metrics do not report how many
configurations were tested before the final one was selected, whether the
observed result is supported by enough history, or whether the apparent
edge survives outside the fitting sample. Each of these issues has a
corresponding treatment in the statistical backtesting literature and
represents a distinct source of estimation or selection bias.

The MinervaScore does not introduce new statistical corrections for these
effects; it aggregates existing corrections and reports them through a
verdict-consistent display. Its additional component is the regime-stability
diagnostic, which is treated separately from the established validation
quantities.

The MinervaScore is therefore distinct from a search objective such as the
GT-Score, which is a common practitioner benchmark. The GT-Score is
an empirical figure of merit based on a combination of equity-curve
descriptors and is optimized during the search process. The MinervaScore is instead applied after selection as a validation layer
and is primarily built from validation quantities with explicit
admissibility thresholds. The difference is conceptual rather than a direct
comparison of predictive accuracy: the two approaches discriminate genuine
edge at similar levels in the synthetic experiments
(\S\ref{sec:synthetic}). The contribution of the statistical formulation is
therefore not a larger score, but the ability to account for multiple
testing, separate different failure modes, and audit the individual
components.

In the ground-truth study, the GT-Score proxy performs similarly on easier
problems but loses discrimination in more difficult regimes, reaching a
worst-case AUROC of 0.806, where selection effects and overfitting have a
larger impact on observed performance. The MinervaScore explicitly accounts
for these effects through its validation gates and only assigns
certification when all five conditions are satisfied. It is not a predictor
of future profitability: the sealed real-market study did not demonstrate forward predictive power
on the tested population, which showed little detectable edge. The purpose
of the MinervaScore is instead to measure how much of an observed backtest
result remains after accounting for known sources of statistical
overstatement.

\subsection{Summary and future work}

The MinervaScore, an aggregation and presentation layer that integrates a five-gate backtest-validation battery into a single verdict-consistent score, was introduced and assessed in this paper. The construction rests on two design choices and one empirical observation. A number of gate outputs are empirically shown to be substantially concentrated at boundary values in production populations; for instance, 60--95\% of DSR values are precisely zero. This loss of information is thus inherited by a composite that is built directly from the raw $[0,1]$ coordinates. The ordering eliminated by the raw representation is restored by expressing each margin on a scale where the accompanying gate stays informative---the logit transformation for bounded gates and the pre-$\Phi$ statistic for the DSR.
By construction, a displayed score is at least 80 if and only if all five gates pass. This property holds true for all 359{,}062 production records. The second design choice is that unavailable inputs are treated cautiously; instead of being substituted by an implicit approximation, missing information inhibits scoring.

These design decisions are supported by the empirical findings. On the
pre-specified difficulty grid, the MinervaScore is the most stable
composite measure (worst-case regret 0.021 compared with 0.071--0.138).
At the headline difficulty, it scores slightly above the GT-Score proxy,
with a positive bootstrap interval and an AUROC of 0.989 in controlled
ground-truth testing. After correcting the null benchmark, however, the
corrected DSR-alone baseline accounts for most of the discriminative
performance and remains the strongest single-signal baseline in parts of
the grid. The contribution of the full composite is therefore not a large
increase in discrimination, but broader coverage across failure modes.

The individual components also respond to search size and history length in
ways consistent with their statistical interpretation. In addition, the
evaluation framework rejected a proposed DSR modification that increased
scores without improving separation, while supporting the Lo-null benchmark
correction through ground-truth evidence. The sealed real-market study was
reported separately and returned a null result, with no evidence of a
forward predictive relationship in the tested population.

The limitations described in \S\ref{sec:discussion} constrain the scope of
the conclusions. The MinervaScore is a ranking, not a calibrated probability,
and no forward predictive power was demonstrated on the real-market sample
studied here. Within these limits, the layer provides a consistent way to
combine multiple validation criteria while preserving the underlying
verdict, avoiding score saturation, and maintaining an interpretable
ordering of strategies according to their supporting evidence.

These findings lead to four directions. In order to assess any potential correlation between score and future outcomes, the real-market study should first be replicated on a population with quantifiable forward edge.
A natural source for such a rolling forward evaluation is the platform's paper-trading records, where scored strategies automatically accumulate subsequent live performance.

Second, the score can be calibrated in accordance with appropriate scoring rules \citep{gneiting2007scoring}, and the aggregation weights and offset can be inferred from observed performance once sufficient forward outcomes are provided. A data-driven calibration would take the place of the existing audited nominal-to-effective weight study.

Third, the choice of reference population remains an open design question. The constants reported here are calibrated on the platform's full optimizer ledger, so each score is evaluated relative to the complete set of strategies the platform has tested. An alternative is to calibrate within each signal family, so that a strategy is compared only with its family peers.

These two approaches answer different questions. A universal calibration makes scores comparable across families, but can reduce resolution within a family when its margins are tightly clustered and therefore contribute little spread to the pooled dispersion. A per-family calibration preserves more resolution within each family, but scores from different families are no longer directly comparable. Neither approach is uniformly preferable. The appropriate choice depends on whether the goal is to rank strategies within a research program or across different programs.

Determining which calibration is appropriate for each use case, and whether both should be reported together, requires evaluating the two approaches on a population with known forward outcomes. We leave this comparison to future work.

Fourth, since user-driven iteration can create selection effects that the validation gates themselves cannot see, the scoring service should be further safeguarded against adaptive querying. The data presented here support the use of the MinervaScore as a transparent validation and display layer until such extensions are available. It is verdict-consistent, avoids saturation, and offers a quantitative ordering of strategies while clearly stating its assumptions and limits.

\section*{Reproducibility}
The reference implementation is the production scoring module at the frozen
calibration commit; the two-scale margin construction of this paper is the
module's shipped default. All production data --- the trial-ledger
population that froze the constants (\S\ref{sec:calibration}) and the
real-market optimizations and sealed-window backtests of
\S\ref{sec:realmarket} --- were produced end-to-end on the Minerva
platform (\url{https://minerva1.com}), the production system that ships
the scoring module documented here. The calibration constants and the
audit (\S\ref{sec:calibration}, \S\ref{sec:audit}) are derived by a
released script run against the platform's trial ledger (de-identified
gate values, bar counts, and per-row coverage metadata; no account,
instrument, or return-series identifiers); the frozen artifact records the
adopted derivation alongside both rejected full-refit alternatives, and
the prior vintage's artifacts are preserved next to the current ones. The
percentile display of \S\ref{sec:display} ships in the same module (81
frozen quantile knots at 1.25\% steps, certification floor at 80), with
serve-time recomputation on every public result surface and drift tests
pinning the knots. The synthetic study and the
component-behavior sweeps (\S\ref{sec:synthetic}) are self-contained
scripts that import the production gate functions and generate their own
labeled strategies from fixed master seeds. The real-market run manifests, enrollment rules,
exclusion rules, and analysis plan are committed to the repository before
execution, and the batch export carries a cryptographic attestation
(row-level SHA-256) of the ledger it was drawn from. Every number in every
table and figure is produced by these scripts; none are typed by hand.

\bibliographystyle{plainnat}
\bibliography{refs}

@article{arian2024backtest,
  author  = {Arian, Hamid R. and Norouzi Mobarekeh, Daniel and Seco, Luis A.},
  title   = {Backtest overfitting in the machine learning era: A comparison of out-of-sample testing methods in a synthetic controlled environment},
  journal = {Knowledge-Based Systems},
  volume  = {305},
  pages   = {112477},
  year    = {2024},
  doi     = {10.1016/j.knosys.2024.112477}
}

@article{bailey2012sharpe,
  author  = {Bailey, David H. and L{\'o}pez de Prado, Marcos},
  title   = {The Sharpe ratio efficient frontier},
  journal = {The Journal of Risk},
  volume  = {15},
  number  = {2},
  pages   = {3--44},
  year    = {2012},
  doi     = {10.21314/JOR.2012.255},
  note    = {Source of the Minimum Track Record Length (MinTRL)}
}

@article{bailey2014dsr,
  author  = {Bailey, David H. and L{\'o}pez de Prado, Marcos},
  title   = {The deflated Sharpe ratio: Correcting for selection bias, backtest overfitting, and non-normality},
  journal = {The Journal of Portfolio Management},
  volume  = {40},
  number  = {5},
  pages   = {94--107},
  year    = {2014},
  doi     = {10.3905/jpm.2014.40.5.094}
}

@article{bailey2017pbo,
  author  = {Bailey, David H. and Borwein, Jonathan M. and L{\'o}pez de Prado, Marcos and Zhu, Qiji J.},
  title   = {The probability of backtest overfitting},
  journal = {Journal of Computational Finance},
  volume  = {20},
  number  = {4},
  pages   = {39--69},
  year    = {2017},
  doi     = {10.21314/JCF.2016.322}
}

@article{benjamini1995fdr,
  author  = {Benjamini, Yoav and Hochberg, Yosef},
  title   = {Controlling the false discovery rate: A practical and powerful approach to multiple testing},
  journal = {Journal of the Royal Statistical Society: Series B},
  volume  = {57},
  number  = {1},
  pages   = {289--300},
  year    = {1995}
}

@article{benjamini2001fdr,
  author  = {Benjamini, Yoav and Yekutieli, Daniel},
  title   = {The control of the false discovery rate in multiple testing under dependency},
  journal = {The Annals of Statistics},
  volume  = {29},
  number  = {4},
  pages   = {1165--1188},
  year    = {2001},
  doi     = {10.1214/aos/1013699998}
}

@article{brown1975method,
  author  = {Brown, Morton B.},
  title   = {A method for combining non-independent, one-sided tests of significance},
  journal = {Biometrics},
  volume  = {31},
  number  = {4},
  pages   = {987--992},
  year    = {1975},
  doi     = {10.2307/2529826}
}

@article{cinar2022poolr,
  author  = {{\c{C}}{\i}nar, Ozan and Viechtbauer, Wolfgang},
  title   = {The poolr package for combining independent and dependent p-values},
  journal = {Journal of Statistical Software},
  volume  = {101},
  number  = {1},
  pages   = {1--42},
  year    = {2022},
  doi     = {10.18637/jss.v101.i01}
}

@article{cont2001stylized,
  author  = {Cont, Rama},
  title   = {Empirical properties of asset returns: Stylized facts and statistical issues},
  journal = {Quantitative Finance},
  volume  = {1},
  number  = {2},
  pages   = {223--236},
  year    = {2001},
  doi     = {10.1080/713665670}
}

@article{dwork2015reusable,
  author  = {Dwork, Cynthia and Feldman, Vitaly and Hardt, Moritz and Pitassi, Toniann and Reingold, Omer and Roth, Aaron},
  title   = {The reusable holdout: Preserving validity in adaptive data analysis},
  journal = {Science},
  volume  = {349},
  number  = {6248},
  pages   = {636--638},
  year    = {2015},
  doi     = {10.1126/science.aaa9375}
}

@article{gneiting2007scoring,
  author  = {Gneiting, Tilmann and Raftery, Adrian E.},
  title   = {Strictly proper scoring rules, prediction, and estimation},
  journal = {Journal of the American Statistical Association},
  volume  = {102},
  number  = {477},
  pages   = {359--378},
  year    = {2007},
  doi     = {10.1198/016214506000001437}
}

@article{hansen2011mcs,
  author  = {Hansen, Peter R. and Lunde, Asger and Nason, James M.},
  title   = {The model confidence set},
  journal = {Econometrica},
  volume  = {79},
  number  = {2},
  pages   = {453--497},
  year    = {2011},
  doi     = {10.3982/ECTA5771}
}

@article{hansen2005spa,
  author  = {Hansen, Peter Reinhard},
  title   = {A test for superior predictive ability},
  journal = {Journal of Business and Economic Statistics},
  volume  = {23},
  number  = {4},
  pages   = {365--380},
  year    = {2005},
  doi     = {10.1198/073500105000000063}
}

@article{hartung1999note,
  author  = {Hartung, Joachim},
  title   = {A note on combining dependent tests of significance},
  journal = {Biometrical Journal},
  volume  = {41},
  number  = {7},
  pages   = {849--855},
  year    = {1999}
}

@article{harvey2015backtesting,
  author  = {Harvey, Campbell R. and Liu, Yan},
  title   = {Backtesting},
  journal = {The Journal of Portfolio Management},
  volume  = {42},
  number  = {1},
  pages   = {13--28},
  year    = {2015},
  doi     = {10.3905/jpm.2015.42.1.013},
  note    = {The working-paper version (2014) introduces the haircut Sharpe}
}

@article{harvey2016cross,
  author  = {Harvey, Campbell R. and Liu, Yan and Zhu, Heqing},
  title   = {\ldots and the cross-section of expected returns},
  journal = {The Review of Financial Studies},
  volume  = {29},
  number  = {1},
  pages   = {5--68},
  year    = {2016},
  doi     = {10.1093/rfs/hhv059}
}

@incollection{liptak1958combination,
  author    = {Lipt{\'a}k, Tam{\'a}s},
  title     = {On the combination of independent tests},
  booktitle = {Magyar Tudom{\'a}nyos Akad{\'e}mia Matematikai Kutat{\'o} Int{\'e}zet{\'e}nek K{\"o}zlem{\'e}nyei},
  volume    = {3},
  pages     = {171--197},
  publisher = {Hungarian Academy of Sciences},
  year      = {1958},
  note      = {Origin of the weighted inverse-normal (weighted-Z) combination}
}

@article{lo2002sharpe,
  author  = {Lo, Andrew W.},
  title   = {The statistics of Sharpe ratios},
  journal = {Financial Analysts Journal},
  volume  = {58},
  number  = {4},
  pages   = {36--52},
  year    = {2002},
  doi     = {10.2469/faj.v58.n4.2453}
}

@book{lopezdeprado2018afml,
  author    = {L{\'o}pez de Prado, Marcos},
  title     = {Advances in Financial Machine Learning},
  publisher = {Wiley},
  address   = {Hoboken, NJ},
  year      = {2018},
  isbn      = {978-1119482086}
}

@misc{manheim2018goodhart,
  author = {Manheim, David and Garrabrant, Scott},
  title  = {Categorizing variants of Goodhart's law},
  year   = {2018},
  howpublished = {arXiv preprint},
  url    = {https://arxiv.org/abs/1803.04585}
}

@article{mcshane2019abandon,
  author  = {McShane, Blakeley B. and Gal, David and Gelman, Andrew and Robert, Christian and Tackett, Jennifer L.},
  title   = {Abandon statistical significance},
  journal = {The American Statistician},
  volume  = {73},
  number  = {sup1},
  pages   = {235--245},
  year    = {2019},
  doi     = {10.1080/00031305.2018.1527253}
}

@techreport{mertens2002comments,
  author      = {Mertens, Elmar},
  title       = {Comments on variance of the IID estimator in Lo (2002)},
  institution = {University of Basel},
  type        = {Working paper},
  year        = {2002},
  note        = {Non-normal (skew/kurtosis) correction to the Sharpe-ratio estimator variance; the i.i.d.\ adjustment used by the DSR}
}

@article{newey1987simple,
  author  = {Newey, Whitney K. and West, Kenneth D.},
  title   = {A simple, positive semi-definite, heteroskedasticity and autocorrelation consistent covariance matrix},
  journal = {Econometrica},
  volume  = {55},
  number  = {3},
  pages   = {703--708},
  year    = {1987},
  doi     = {10.2307/1913610}
}

@article{paruolo2013ratings,
  author  = {Paruolo, Paolo and Saisana, Michaela and Saltelli, Andrea},
  title   = {Ratings and rankings: Voodoo or science?},
  journal = {Journal of the Royal Statistical Society: Series A},
  volume  = {176},
  number  = {3},
  pages   = {609--634},
  year    = {2013},
  doi     = {10.1111/j.1467-985X.2012.01059.x}
}

@incollection{platt1999probabilistic,
  author    = {Platt, John C.},
  title     = {Probabilistic outputs for support vector machines and comparisons to regularized likelihood methods},
  booktitle = {Advances in Large Margin Classifiers},
  pages     = {61--74},
  publisher = {MIT Press},
  year      = {1999}
}

@article{poole2016combining,
  author  = {Poole, William and Gibbs, David L. and Shmulevich, Ilya and Bernard, Brady and Knijnenburg, Theo A.},
  title   = {Combining dependent {P}-values with an empirical adaptation of Brown's method},
  journal = {Bioinformatics},
  volume  = {32},
  number  = {17},
  pages   = {i430--i436},
  year    = {2016},
  doi     = {10.1093/bioinformatics/btw438}
}

@article{romano2005stepwise,
  author  = {Romano, Joseph P. and Wolf, Michael},
  title   = {Stepwise multiple testing as formalized data snooping},
  journal = {Econometrica},
  volume  = {73},
  number  = {4},
  pages   = {1237--1282},
  year    = {2005},
  doi     = {10.1111/j.1468-0262.2005.00615.x}
}

@article{saisana2005uncertainty,
  author  = {Saisana, Michaela and Saltelli, Andrea and Tarantola, Stefano},
  title   = {Uncertainty and sensitivity analysis techniques as tools for the quality assessment of composite indicators},
  journal = {Journal of the Royal Statistical Society: Series A},
  volume  = {168},
  number  = {2},
  pages   = {307--323},
  year    = {2005},
  doi     = {10.1111/j.1467-985X.2005.00350.x}
}

@article{sheppert2026gtscore,
  author  = {Sheppert, Alexander Pearson},
  title   = {The {GT}-score: A robust objective function for reducing overfitting in data-driven trading strategies},
  journal = {Journal of Risk and Financial Management},
  volume  = {19},
  number  = {1},
  pages   = {60},
  year    = {2026},
  doi     = {10.3390/jrfm19010060},
  note    = {Also arXiv:2602.00080}
}

@book{stouffer1949american,
  author    = {Stouffer, Samuel A. and Suchman, Edward A. and DeVinney, Leland C. and Star, Shirley A. and Williams, Robin M.},
  title     = {The American Soldier, Vol.~1: Adjustment During Army Life},
  publisher = {Princeton University Press},
  address   = {Princeton, NJ},
  year      = {1949},
  note      = {Unweighted inverse-normal (Z) combination}
}

@article{wasserstein2016asa,
  author  = {Wasserstein, Ronald L. and Lazar, Nicole A.},
  title   = {The {ASA} statement on p-values: Context, process, and purpose},
  journal = {The American Statistician},
  volume  = {70},
  number  = {2},
  pages   = {129--133},
  year    = {2016},
  doi     = {10.1080/00031305.2016.1154108}
}

@article{white2000reality,
  author  = {White, Halbert},
  title   = {A reality check for data snooping},
  journal = {Econometrica},
  volume  = {68},
  number  = {5},
  pages   = {1097--1126},
  year    = {2000},
  doi     = {10.1111/1468-0262.00152}
}

@article{whitlock2005combining,
  author  = {Whitlock, Michael C.},
  title   = {Combining probability from independent tests: The weighted {Z}-method is superior to Fisher's approach},
  journal = {Journal of Evolutionary Biology},
  volume  = {18},
  number  = {5},
  pages   = {1368--1373},
  year    = {2005},
  doi     = {10.1111/j.1420-9101.2005.00917.x}
}

@misc{yin2026implementation,
  author = {Yin, Don and Miki, Takeshi and Lesnichenko, Vladislav and Gural, Vasyl},
  title  = {Implementation risk in portfolio backtesting: A previously unquantified source of error},
  year   = {2026},
  howpublished = {arXiv preprint},
  url    = {https://arxiv.org/abs/2603.20319}
}

@article{hanley1982roc,
  author  = {Hanley, James A. and McNeil, Barbara J.},
  title   = {The meaning and use of the area under a receiver operating characteristic ({ROC}) curve},
  journal = {Radiology},
  volume  = {143},
  number  = {1},
  pages   = {29--36},
  year    = {1982},
  doi     = {10.1148/radiology.143.1.7063747}
}

@article{fawcett2006roc,
  author  = {Fawcett, Tom},
  title   = {An introduction to {ROC} analysis},
  journal = {Pattern Recognition Letters},
  volume  = {27},
  number  = {8},
  pages   = {861--874},
  year    = {2006},
  doi     = {10.1016/j.patrec.2005.10.010}
}

@book{holland1975adaptation,
  author    = {Holland, John H.},
  title     = {Adaptation in Natural and Artificial Systems},
  publisher = {University of Michigan Press},
  address   = {Ann Arbor},
  year      = {1975}
}

@book{goldberg1989genetic,
  author    = {Goldberg, David E.},
  title     = {Genetic Algorithms in Search, Optimization, and Machine Learning},
  publisher = {Addison-Wesley},
  address   = {Reading, MA},
  year      = {1989}
}

@article{miller1995genetic,
  author  = {Miller, Brad L. and Goldberg, David E.},
  title   = {Genetic algorithms, tournament selection, and the effects of noise},
  journal = {Complex Systems},
  volume  = {9},
  number  = {3},
  pages   = {193--212},
  year    = {1995}
}

\end{document}